\RequirePackage{ifpdf}
\ifpdf 
\documentclass[pdftex]{sigma}
\else
\documentclass{sigma}
\fi

\numberwithin{equation}{section}

\begin{document}

\allowdisplaybreaks

\renewcommand{\PaperNumber}{107}

\FirstPageHeading

\ShortArticleName{Properties of the Exceptional ($X_{\ell}$) Laguerre and Jacobi Polynomials}

\ArticleName{Properties of the Exceptional ($\boldsymbol{X_{\ell}}$) Laguerre \\ and Jacobi Polynomials}

\Author{Choon-Lin HO~$^\dag$, Satoru ODAKE~$^\ddag$ and Ryu SASAKI~$^\S$}

\AuthorNameForHeading{C.-L.~Ho, S.~Odake and R.~Sasaki}

\Address{$^\dag$~Department of Physics, Tamkang University,
    Tamsui 251, Taiwan (R.O.C.)}
\EmailD{\href{mailto:hcl@mail.tku.edu.tw}{hcl@mail.tku.edu.tw}}

\Address{$^\ddag$~Department of Physics, Shinshu University,
    Matsumoto 390-8621, Japan}
\EmailD{\href{mailto:odake@azusa.shinshu-u.ac.jp}{odake@azusa.shinshu-u.ac.jp}}

\Address{$^\S$~Yukawa Institute for Theoretical Physics,
    Kyoto University, Kyoto 606-8502, Japan}
    \EmailD{\href{mailto:ryu@yukawa.kyoto-u.ac.jp}{ryu@yukawa.kyoto-u.ac.jp}}

\ArticleDates{Received April 18, 2011, in f\/inal form November 19, 2011;  Published online November 25, 2011}

\Abstract{We present various results on the properties of the
four inf\/inite sets of the excep\-tional $X_{\ell}$ polynomials
discovered recently by  Odake and  Sasaki [{\it Phys. Lett.~B} {\bf 679}
(2009), 414--417; {\it Phys. Lett.~B} {\bf 684} (2010), 173--176]. These $X_{\ell}$
polynomials are global solutions of second order Fuchsian
dif\/ferential equations with $\ell+3$ regular singularities and
their conf\/luent limits. We derive equivalent but much simpler
looking forms of the $X_{\ell}$ polynomials. The other subjects
discussed in detail are: factorisation of the Fuchsian
dif\/ferential operators, shape invariance, the forward and backward
shift operations, invariant polynomial subspaces under the
Fuchsian dif\/ferential operators, the Gram--Schmidt
orthonormalisation procedure, three term recurrence relations and
the generating functions for the $X_{\ell}$ polynomials.}


\Classification{42C05; 33E30; 81Q05}

\newcommand{\eqdef}{\stackrel{\text{def}}{=}}
\newcommand{\n}{\nonumber\\}
\newcommand{\bm}{\boldsymbol}
\newcommand{\ignore}[1]{}
\newcommand{\Xil}[1]{\Xi_{\ell,\bm{\lambda}}[#1]}
\newcommand{\Xilbig}[1]{\Xi_{\ell,\bm{\lambda}}\bigl[#1\bigr]}
\newcommand{\cF}{c_{\text{\tiny$\mathcal{F}$}}}

\section{Introduction}
\label{sec:intro}

\looseness=1
Four sets of inf\/initely many exceptional ($X_{\ell}$) polynomials
satisfying second order dif\/ferential equations were introduced recently
by two of the present authors \cite{os16, os19}.
They were obtained as the main part of the eigenfunctions of exactly
solvable one-dimensional quantum mechanical systems which were
deformations of the well-known solvable systems of the radial oscillator
\cite{infhul,susyqm} and the trigonometric Darboux--P\"{o}schl--Teller (DPT)
potential \cite{dpt} by a degree $\ell$ eigenpolynomial.
Thus the orthogonality and completeness of the~$X_{\ell}$ polynomials are
automatically guaranteed. These polynomials, termed exceptional Laguerre
and Jacobi polynomials, have two types in each family, L1, L2 and J1 and J2.
The Laguerre family L1~(L2) is obtained from the Jacobi family~J1~(J2)
by the well-known limit~\eqref{J->L}, which takes the Jacobi poly\-no\-mials
to the Laguerre polynomials. The J1 and J2 are mirror images of each other,
see~\eqref{Jparity}, but their limits L1 and L2 are clearly distinct.
These polynomials are {\em exceptional\/} in the sense that they start
at degree $\ell$ ($\ell=1,2,\ldots$) rather than degree 0 constant term.
Thus they are not constrained by Bochner's theorem \cite{bochner},
which states that the orthogonal polynomials (star\-ting with degree~0)
satisfying a second order dif\/ferential equations are very limited.
Namely, they are only the {\em classical orthogonal polynomials\/},
the Hermite, Laguerre, Jacobi and Bessel polynomials.

\looseness=1
The concept of exceptional orthogonal polynomials was introduced in 2008
by Gomez-Ullate et al.~\cite{gomez1, gomez2}. Within the Sturm--Liouville
theory they constructed $X_1$ Laguerre and $X_1$ Jacobi polynomials,
which turned out to be the f\/irst members of the inf\/inite families.
The results in~\cite{gomez1, gomez2} were reformulated in the framework
of quantum mechanics and shape-invariant potentials~\cite{genden} by
Quesne and collaborators~\cite{quesne,BQR}. They found the f\/irst member
of the deformed hyperbolic DPT potential family, which was also given
in~\cite{os16}.
Quantum mechanical reformulation of\/fers two merits. Firstly, the
orthogonality and completeness of the obtained  eigenfunctions are
guaranteed. Secondly, the well established solution mechanism of shape
invariance combined with Crum's method~\cite{crum}, or the so-called
factorisation method \cite{infhul} or the susy quantum mechanics~\cite{susyqm} is available.
For systems consisting of discrete eigenvalues only,
shape invariance is a well-known suf\/f\/icient
condition for exact solvability of one-dimensional Schr\"{o}dinger equation.
The discovery of the four sets of inf\/initely many exceptional orthogonal
polynomials was achieved by pursuing shape invariant deformation~\cite{os16, os19}.
To be more precise, deformation of the potential by keeping the property
of shape invariance.
After the f\/irst paper on inf\/initely many $X_{\ell}$
polynomials~\cite{os16}, Quesne reported a type II $X_2$ Laguerre
polynomials~\cite{quesne2,tanaka}, which led to the discovery of the
L2 family of $X_{\ell}$ Laguerre polynomials~\cite{os19}.
In a previous paper~\cite{os18}, two of the present authors unveiled
inf\/initely many polynomial identities of degree $3\ell$ involving cubic
products of the Laguerre or the Jacobi polynomials, which encode the
information of exact solvability of the dif\/ferential equations
governing the $X_{\ell}$ poly\-nomials.

In this paper we explore various properties of the $X_{\ell}$ polynomials.
We emphasise that the~J1 and~J2 polynomials are the global solutions of
a Fuchsian dif\/ferential equation having $\ell+3$ regular singularities.
They are located at $\pm1,\infty$ and the $\ell$ zeros of the polynomial
$\xi_{\ell}(\eta;\bm{\lambda})$~\eqref{defxiL},~\eqref{defxiJ} which is
used for the deformation.
Factorisation and shape invariance are reformulated accordingly, leading
to Rodrigues formulas and the forward and backward shift operations.
The existence of the extra regular singularities implies that the
ordinary vector space spanned by degree $n$ polynomials
$\mathcal{V}_{0,n}=\,\text{Span}\,[1,x,\ldots,x^n]$ is not invariant
under the Fuchsian dif\/ferential operator~\eqref{vnotinv}. Appropriate
invariant polynomial subspaces are introduced and their properties are
used to derive the explicit forms of the exceptional polynomials.
Some of the important subjects in orthogonal polynomial theory, namely,
the Gram--Schmidt orthonormalisation, the generating functions,
three term recurrence relations,
the zeros of these orthogonal polynomials, etc, are also discussed.
New inf\/initely many polynomial identities underlying the forward and
backward shift operations are also reported.

The plan of this paper is as follows. In Section~\ref{section2} we present the
explicit forms of the four sets of inf\/initely many exceptional
orthogonal polynomials together with their weight functions and
the normalisation constants. They are equal to those reported
earlier \cite{os16, os19}, but look much simpler than the original
ones. The new forms of the polynomials reveal the structure of the
theory. In Section~\ref{section3} the Fuchsian dif\/ferential equations
governing these polynomials are discussed. Shape invariance and
Rodrigues formulas are presented in Section~\ref{section4}. The identities
underlying the forward and backward shift relations are presented
in Section~\ref{section5}. The polynomial subspaces invariant under the
Fuchsian dif\/ferential operator are discussed in Section~\ref{section6}. This
gives another concise proof of the new forms of the exceptional
polynomials. In Section~\ref{section7} we provide the integration formula
which is essential for the Gram--Schmidt construction in
Section~\ref{section8}. Section~\ref{section9} gives the generating functions for the
$X_{\ell}$ polynomials. The double generating function, that is,
the generating function of the generating functions, is presented
for the L1 and L2 exceptional Laguerre polynomials. A substitute
of the three term recurrence relations for the $X_\ell$
polynomials is introduced in Section~\ref{section10}. In Section~\ref{section11} we state
the qualitative features of the extra zeros of the $X_{\ell}$
polynomials without proof. The f\/inal section is for a summary and
comments. Some technical details are relegated to Appendices. The
equality of the new and original forms of the $X_{\ell}$
polynomials is demonstrated in Appendix~\ref{appendixA}. Forward and backward shift
relations are proved in Appendix~\ref{appendixB}. Derivation of the integration formula
is provided in Appendix~\ref{appendixC}. The properties of the $X_\ell$ Jacobi
polynomials as the solutions of the quantum mechanical systems
with deformed hyperbolic DPT potentials are summarised in Appendix~\ref{appendixD}. A
concise summary of some important properties of the Laguerre and
Jacobi polynomials is given in Appendix~\ref{appendixE} for self-containedness.

Throughout this paper we stick to the notation of our previous
papers~\cite{os16, os19, os18, os17}. Reference to the quantum
mechanical language is made minimal in order to attain wider
readership than before. Most concepts and formulas are common to
the four sets of exceptional polynomials. As far as possible we
use generic formulas valid for  all the four dif\/ferent sets of
$X_{\ell}$ polynomials, in order to emphasise the underlying
structure and at the same time to avoid redundancy.

\section{Exceptional Laguerre and Jacobi polynomials}\label{section2}

\looseness=-1
Here we present four sets of inf\/initely many exceptional orthogonal
polynomials \cite{os16,os19}, among them two are deformations of the
Laguerre polynomials, and the others are deformations of~the Jacobi
polynomials. They are expressed as a bilinear form of the original
polynomials, the Laguer\-re or Jacobi polynomials and the deforming
polynomials, depending on the set of parame\-ters~$\bm{\lambda}$ and
their shifts $\bm{\delta}$ and a non-negative integer $\ell$, which
is the degree of the defor\-ming po\-ly\-no\-mials.
It is important to stress that these explicit forms \eqref{XLform}--\eqref{defxiJ} can be derived by the Darboux--Crum transformations
starting from the original (Laguerre or Jacobi) polyno\-mials~\mbox{\cite{gkm10,stz}}.
The two sets of {\em exceptional Laguerre polynomials\/}
($\ell=0,1,2,\ldots$, $n=0,1,2,\ldots$) are:
\begin{gather}
 P_{\ell,n}(\eta;\bm{\lambda})\eqdef
 \begin{cases}
 \xi_{\ell}(\eta;\bm{\lambda}+\bm{\delta})P_n(\eta;g+\ell-1)
 -\xi_{\ell}(\eta;\bm{\lambda})\partial_{\eta}
 P_n(\eta;g+\ell-1),\quad & (\text{L1})\vspace{1mm}\\
 \big(n+g+\frac12\big)^{-1}\bigl(\big(g+\frac12\big)
 \xi_{\ell}(\eta;\bm{\lambda}+\bm{\delta})P_n(\eta;g+\ell+1)\\
\hphantom{\big(n+g+\frac12\big)^{-1}\bigl(}{}
 {} +\eta\xi_{\ell}(\eta;\bm{\lambda})\partial_{\eta}P_n(\eta;g+\ell+1)
 \bigr), \quad & (\text{L2})
 \end{cases}
 \label{XLform}
\end{gather}
in which $\bm{\lambda}\eqdef g>0$ and $\bm{\delta}\eqdef 1$ and
\begin{gather}
 P_n(\eta;g)\eqdef L_n^{(g-\frac{1}{2})}(\eta),\qquad
 \xi_{\ell}(\eta;g)\eqdef
 \begin{cases}
 L_{\ell}^{(g+\ell-\frac32)}(-\eta), & (\text{L1})\vspace{1mm}\\
 L_{\ell}^{(-g-\ell-\frac12)}(\eta).\quad & (\text{L2})
 \end{cases}
 \label{defxiL}
\end{gather}
The two sets of {\em exceptional Jacobi polynomials}
($\ell=0,1,2,\ldots$, $n=0,1,2,\ldots$) are\footnote{It should be remarked that the naming J1 and J2 are interchanged from the
previous ones~\cite{os19} in order to respect the logical consistency
rather than hysteresis.}:
\begin{gather}
 P_{\ell,n}(\eta;\bm{\lambda})\eqdef
 \begin{cases}
 \big(n+h+\frac12\big)^{-1}\bigl(
 \big(h+\frac12\big)\xi_{\ell}(\eta;\bm{\lambda}+\bm{\delta})
 P_n(\eta;g+\ell-1,h+\ell+1)\quad &\\
 \phantom{\big(n+h+\frac12\big)^{-1}\bigl(}
 +(1+\eta)\xi_{\ell}(\eta;\bm{\lambda})
 \partial_{\eta}P_{n}(\eta;g+\ell-1,h+\ell+1)\bigr),
 \!\!\!\!\quad & (\text{J1})\vspace{1mm}\\
 \big(n+g+\frac12\big)^{-1}\bigl(
 \big(g+\frac12\big)\xi_{\ell}(\eta;\bm{\lambda}+\bm{\delta})
 P_n(\eta;g+\ell+1,h+\ell-1)\quad &\\
 \phantom{\big(n+g+\frac12\big)^{-1}\bigl(}
 -(1-\eta)\xi_{\ell}(\eta;\bm{\lambda})
 \partial_{\eta}P_{n}(\eta;g+\ell+1,h+\ell-1)\bigr),\!\!\!\!\quad
 & (\text{J2})
 \end{cases} \!\!\!\!\!\!\!\!\!\!\!\!
 \label{XJform}
\end{gather}
in which $\bm{\lambda}\eqdef(g,h)$, $g>0$, $h>0$, $\bm{\delta}\eqdef(1,1)$
and
\begin{gather}
 P_n(\eta;g,h)\eqdef P_n^{(g-\frac12,h-\frac12)}(\eta),\nonumber\\
 \xi_{\ell}(\eta;g,h)\eqdef
 \begin{cases}
 P_{\ell}^{(g+\ell-\frac32,-h-\ell-\frac12)}(\eta),\quad  g>h>0,\quad & (\text{J1}) \\
 P_{\ell}^{(-g-\ell-\frac12,h+\ell-\frac32)}(\eta),\quad h>g>0. \quad  & (\text{J2})
 \end{cases}
 \label{defxiJ}
\end{gather}
It should be stressed that the deforming polynomial
$\xi_\ell(\eta;\bm{\lambda})$ does not have a zero in the domain of
orthogonality, $(0,\infty)$ for~L1 and~L2 and $(-1,1)$ for~J1 and~J2.
This is demonstrated explicitly in~(2.39),~(2.40) of~\cite{os18}.
These $X_{\ell}$ polynomials have the following general structure{\samepage
\begin{gather}
 d_0(n,\bm{\lambda})P_{\ell,n}(\eta;\bm{\lambda})
 =d_1(\bm{\lambda})\xi_{\ell}(\eta;\bm{\lambda}+\bm{\delta})
 P_n(\eta;\bm{\lambda}+\ell\bm{\delta}+\bm{\tilde{\delta}})\nonumber\\
 \hphantom{d_0(n,\bm{\lambda})P_{\ell,n}(\eta;\bm{\lambda})=}{}
 -d_2(\eta)\xi_{\ell}(\eta;\bm{\lambda})
 \partial_{\eta}P_n(\eta;\bm{\lambda}+\ell\bm{\delta}+\bm{\tilde{\delta}}),
 \label{Plnform}
\end{gather}
where}
\begin{gather}
 d_0(n,\bm{\lambda}) \eqdef
 \begin{cases}
 1, \quad & (\text{L1})\\
 n+g+\frac12,\quad & (\text{L2,\ J2})\vspace{1mm}\\
 n+h+\frac12, \quad & (\text{J1})
 \end{cases}
 \qquad
 \bm{\tilde{\delta}}\eqdef
 \begin{cases}
 \mp 1,\quad &  (\text{L1/L2})\\
 \mp(1,-1), \quad & (\text{J1/J2})
 \end{cases}
\nonumber \\
  d_1(\bm{\lambda}) \eqdef
 \begin{cases}
 1, \quad & (\text{L1})\\
 g+\frac12,\quad & (\text{L2,\ J2})\vspace{1mm}\\
 h+\frac12,\quad &  (\text{J1})
 \end{cases}
 \qquad
 d_2(\eta)\eqdef
 \begin{cases}
 1,\quad & (\text{L1})\\
 -\eta,\quad & (\text{L2})\\
 \mp(1\pm \eta).\quad & (\text{J1/J2})
 \end{cases}
 \label{d12def}
\end{gather}
We introduce a linear map $\Xil{\,\cdot\,}$ for a dif\/ferentiable function
$p(\eta)$,
\begin{gather}
 \Xil{p(\eta)}\eqdef
 d_1(\bm{\lambda})\xi_{\ell}(\eta;\bm{\lambda}+\bm{\delta})p(\eta)
 -d_2(\eta)\xi_{\ell}(\eta;\bm{\lambda})\partial_{\eta}p(\eta).
 \label{Xil}
\end{gather}
Then the $X_{\ell}$ polynomial \eqref{Plnform} is expressed succinctly as
\begin{gather}
 d_0(n,\bm{\lambda})P_{\ell,n}(\eta;\bm{\lambda})
 =\Xilbig{P_n(\eta;\bm{\lambda}+\ell\bm{\delta}+\bm{\tilde{\delta}})}.
 \label{Pln=Xil[Pn]}
\end{gather}

The $X_0$ polynomials $P_{0,n}(\eta;\bm{\lambda})=P_n(\eta;\bm{\lambda})$
are the {\em undeformed} polynomials, i.e., the Laguerre or the Jacobi
polynomials themselves. Therefore the above formulas \eqref{XLform},
\eqref{XJform} and \eqref{Plnform} for $\ell=0$ are non-trivial identities
among the Laguerre or the Jacobi polynomials
\begin{gather*}
 d_0(n,\bm{\lambda})P_{n}(\eta;\bm{\lambda})
 =d_1(\bm{\lambda})P_n(\eta;\bm{\lambda}+\bm{\tilde{\delta}})
 -d_2(\eta)\partial_{\eta}P_n(\eta;\bm{\lambda}+\bm{\tilde{\delta}}).
\end{gather*}
This is shown by using \eqref{Lforward}, \eqref{Jforward} and the
following; \eqref{Lid1} for $\text{L1}$,
\eqref{Lid1} and \eqref{Lid2} for $\text{L2}$,
\eqref{Jid1} for $\text{J1}$ and \eqref{Jid1m} for $\text{J2}$.
The $X_{\ell}$ polynomials $P_{\ell,n}(\eta;\bm{\lambda})$ are degree
$\ell+n$ polynomials in~$\eta$ and start at degree $\ell$:
\begin{gather}
 P_{\ell,0}(\eta;\bm{\lambda})=\xi_{\ell}(\eta;\bm{\lambda}+\bm{\delta}).
 \label{degreel}
\end{gather}
They are orthogonal with respect to the weight function
$\mathcal{W}_{\ell}(\eta,\bm{\lambda})$ which is a deformation of the
weight function ${W}(\eta;\bm{\lambda})$ for the Laguerre or Jacobi
polynomials:
\begin{gather}
  \int P_{\ell,n}(\eta;\bm{\lambda})P_{\ell,m}(\eta;\bm{\lambda})
 \mathcal{W}_{\ell}(\eta;\bm{\lambda})d\eta
 =h_{\ell,n}(\bm{\lambda})\delta_{nm},
 \label{orthonormalform}\\
  \mathcal{W}_{\ell}(\eta;\bm{\lambda}) \eqdef
 \frac{{W}(\eta;\bm{\lambda}+\ell\bm{\delta})}
 {\xi_{\ell}(\eta;\bm{\lambda})^2},
 \nonumber\\
  W(\eta;\bm{\lambda})\eqdef
\left\{ \!\!\begin{array}{lll}
 \frac12e^{-\eta}\eta^{g-\frac12},\quad &0<\eta<\infty,\quad  & (\text{L})\vspace{1mm}\\
 \frac{1}{2^{g+h+1}}(1-\eta)^{g-\frac12}(1+\eta)^{h-\frac12},\quad
 &-1<\eta<1. \quad & (\text{J})
 \end{array}\right.
 \label{defweight}
\end{gather}
The normalisation constants are meromorphic functions of the parameters
$g$, $h$ and $\ell$:
\begin{alignat}{3}
 & (\text{L}):\quad&&
 h_n(g)\eqdef\frac{1}{2 n!} \Gamma(n+g+\tfrac12),&
 \label{normL0}\\
 &&& h_{\ell,n}(g)\eqdef
h_n(g+\ell)\times
 \begin{cases}
 \frac{n+g+2\ell-\frac12}{n+g+\ell-\frac12},\quad &(\text{L1})\vspace{1mm}\\
 \frac{n+g+\ell+\frac12}{n+g+\frac12},\quad & (\text{L2})
 \end{cases} &
 \label{normL}\\
& (\text{J}):\quad&&
 h_n(g,h)\eqdef\frac{\Gamma(n+g+\frac12)\Gamma(n+h+\frac12)}
 {2\,n!\,(2n+g+h)\Gamma(n+g+h)}, &
 \label{normJ0}\\
&&&h_{\ell,n}(g,h)\eqdef
  h_n(g+\ell,h+\ell)\times
 \begin{cases}
 \frac{(n+h+\ell+\frac12)(n+g+2\ell-\frac12)}
 {(n+h+\frac12)(n+g+\ell-\frac12)}, \quad & (\text{J1})\vspace{2mm}\\
 \frac{(n+g+\ell+\frac12)(n+h+2\ell-\frac12)}
 {(n+g+\frac12)(n+h+\ell-\frac12)}. \quad & (\text{J2})
 \end{cases} &
 \label{normJ}
\end{alignat}

As stressed in \S~3 of~\cite{os19}, the J1 and J2 polynomials are the
mirror images of each other, in the sense $\eta\leftrightarrow-\eta$
and $g\leftrightarrow h$, as exemplif\/ied by the relation
$\xi_{\ell}^{\text{J2}}(\eta;g,h)=(-1)^\ell\xi_{\ell}^{\text{J1}}(-\eta;h,g)$.
However, they lead to the two dif\/ferent sets of the exceptional Laguerre
polynomials, J1$\to$L1, J2$\to$L2. In terms of the limit formulas
\begin{gather}
 \lim_{\beta\to\infty}P_n^{(\alpha,\pm\beta)}\bigl(1-\tfrac{2x}{\beta}\bigr)
 =L_n^{(\alpha)}(\pm x),
 \label{J->L0}
\end{gather}
it is easy to see the relations $P_{\ell,n}(\eta;\bm{\lambda})$
\eqref{XJform}$\to$\eqref{XLform} together with the deforming polynomials
$\xi_{\ell}(\eta;\bm{\lambda})$ \eqref{defxiJ}$\to$\eqref{defxiL},
the normalisation constants
[\eqref{orthonormalform} with \eqref{normJ}]$\to$[\eqref{orthonormalform}
with \eqref{normL}] and others.
The explicit forms of the $X_{\ell}$ polynomials \eqref{XLform} and
\eqref{XJform} are much simpler than those given in the previous papers
\cite{os16,os19}. In Appendix~\ref{appendixA} we will give simple
demonstration that these apparently dif\/ferent forms of $X_{\ell}$
polynomials are in fact equal.

The basic ingredients of the theory of exceptional orthogonal polynomials
are the base polynomial $P_n$ and the deforming polynomial $\xi_{\ell}$.
They satisfy the second order dif\/ferential equations~\eqref{Ldiffeq} and~\eqref{Jdiffeq}, which can be expressed as
\begin{gather}
  c_2(\eta)\partial_{\eta}^2P_n(\eta;\bm{\lambda})
 +c_1(\eta,\bm{\lambda})\partial_{\eta}P_n(\eta;\bm{\lambda})
 =-\tfrac14\mathcal{E}_n(\bm{\lambda})P_n(\eta;\bm{\lambda}),
 \label{Pndiffeq}\\
c_2(\eta)\partial_{\eta}^2\xi_{\ell}(\eta;\bm{\lambda})
 +\tilde{c}_1(\eta,\bm{\lambda},\ell)
 \partial_{\eta}\xi_{\ell}(\eta;\bm{\lambda})
 =-\tfrac14\widetilde{\mathcal{E}}_{\ell}(\bm{\lambda})
 \xi_{\ell}(\eta;\bm{\lambda}),
 \label{xildiffeq}
\end{gather}
where
\begin{gather}
  c_1(\eta,\bm{\lambda})\eqdef
 \begin{cases}
 g+\tfrac12-\eta, \quad & (\text{L})\\
 h-g-(g+h+1)\eta, \quad & (\text{J})
 \end{cases}
 \qquad
 c_2(\eta)\eqdef
 \begin{cases}
 \eta,\quad & (\text{L})\\
 1-\eta^2,\quad & (\text{J})
 \end{cases}
\nonumber \\
\tilde{c}_1(\eta,\bm{\lambda},\ell)\eqdef
 \begin{cases}
 \pm(g+\ell-\tfrac12+\eta),\quad & (\text{L1/L2})\\
 \mp\bigl(g+h+2\ell-1+(g-h)\eta\bigr),\quad & (\text{J1/J2})
 \end{cases}
\nonumber \\
\mathcal{E}_n(\bm{\lambda})\eqdef
 \begin{cases}
 4n,\quad & (\text{L})\\
 4n(n+g+h),\quad & (\text{J})
 \end{cases}
 \qquad
 \widetilde{\mathcal{E}}_{\ell}(\bm{\lambda})\eqdef
 \begin{cases}
 \mp 4\ell,\quad & (\text{L1/L2})\\
 4\ell(\ell\pm g\mp h-1).\quad & (\text{J1/J2})
 \end{cases} \!\!\!\!
 \label{En,Etl}
\end{gather}
The deforming polynomial $\xi_{\ell}(\eta;\bm{\lambda}+\bm{\delta})$
is expressed in terms of $\xi_{\ell}(\eta;\bm{\lambda})$,
\begin{gather}
 d_1(\bm{\lambda})\xi_{\ell}(\eta;\bm{\lambda}+\bm{\delta})
 =d_1(\bm{\lambda}+\ell\bm{\delta})\xi_{\ell}(\eta;\bm{\lambda})
 +d_2(\eta)\partial_{\eta}\xi_{\ell}(\eta;\bm{\lambda}),
 \label{xil(l+d)}
\end{gather}
with $d_1$ and $d_2$ def\/ined in \eqref{d12def}.
This is shown by \eqref{Lforward}, \eqref{Jforward} and various identities of
the polynomials; \eqref{Lid1} for $\text{L1}$,
\eqref{Lid1} and \eqref{Lid2} for $\text{L2}$,
\eqref{Jid1} for $\text{J1}$ and \eqref{Jid1m} for $\text{J2}$.
Conversely $\xi_{\ell}(\eta;\bm{\lambda})$ is expressed in terms of
$\xi_{\ell}(\eta;\bm{\lambda}+\bm{\delta})$,
\begin{gather}
 d_3(\bm{\lambda}+\ell\bm{\delta},\ell)\xi_{\ell}(\eta;\bm{\lambda})
 =d_3(\bm{\lambda},\ell)\xi_{\ell}(\eta;\bm{\lambda}+\bm{\delta})
 +\frac{c_2(\eta)}{d_2(\eta)}\,
 \partial_{\eta}\xi_{\ell}(\eta;\bm{\lambda}+\bm{\delta}),
 \label{xil(l)}
\end{gather}
where
\begin{gather}
 d_3(\bm{\lambda},\ell)\eqdef
 \begin{cases}
 g+\ell-\frac12,\quad & (\text{L1,~J1})\\
 1,\quad & (\text{L2})\\
 h+\ell-\frac12.\quad & (\text{J2})
 \end{cases}
\end{gather}
This is shown in similar ways as above.
The Laguerre and Jacobi dif\/ferential equations for~$P_n$ \eqref{Pndiffeq}
can be factorised into the forward and backward shift relations for $P_n$:
\begin{gather}
  \cF\partial_{\eta}P_n(\eta;\bm{\lambda})
 =f_n(\bm{\lambda})P_{n-1}(\eta;\bm{\lambda}+\bm{\delta}),
 \label{Pnforward}\\
c_1(\eta,\bm{\lambda})P_{n-1}(\eta;\bm{\lambda}+\bm{\delta})
 +c_2(\eta)\partial_{\eta}P_{n-1}(\eta;\bm{\lambda}+\bm{\delta})
 =-\tfrac14\cF b_{n-1}(\bm{\lambda})P_n(\eta;\bm{\lambda}),
 \label{Pnbackward}
\end{gather}
where
\begin{gather}
 \cF\eqdef
 \begin{cases}
 2,\quad & (\text{L})\\
 -4, \quad & (\text{J})
 \end{cases}
 \qquad
 f_n(\bm{\lambda})=
 \begin{cases}
 -2,\quad & (\text{L})\\
 -2(n+g+h),\quad & (\text{J})
 \end{cases}
 \qquad
 b_{n-1}(\bm{\lambda})=-2n.
 \label{CFfnbn-1}
\end{gather}
See \eqref{Lforward}, \eqref{Lbackward} and
\eqref{Jforward}, \eqref{Jbackward} for the explicit forms of the forward
and backward shift relations.
Note that $f_n(\bm{\lambda})$ and $b_{n-1}(\bm{\lambda})$ are the factors
of the eigenvalue
\begin{gather*}
 \mathcal{E}_n(\bm{\lambda})=f_n(\bm{\lambda})b_{n-1}(\bm{\lambda}),
 \qquad n=0,1,\ldots.
\end{gather*}

\section[Fuchsian differential equations with extra $\ell$ regular
singularities]{Fuchsian dif\/ferential equations\\ with extra $\boldsymbol{\ell}$ regular
singularities}\label{section3}

The exceptional Laguerre and Jacobi polynomials satisfy a second order
linear dif\/ferential equation in the {\em entire complex $\eta$ plane\/}:
\begin{gather}
 \widetilde{\mathcal{H}}_{\ell}(\bm{\lambda})
 P_{\ell,n}(\eta;\bm{\lambda})
 =\mathcal{E}_{\ell,n}(\bm{\lambda})P_{\ell,n}(\eta;\bm{\lambda}),\qquad
 \mathcal{E}_{\ell,n}(\bm{\lambda})
 =\mathcal{E}_n(\bm{\lambda}+\ell\bm{\delta}),
 \label{htildeeq}
\end{gather}
in which the eigenvalue $\mathcal{E}_n$ is def\/ined in \eqref{En,Etl}.
The $X_{\ell}$ polynomials are not constrained by Bochner's theorem~\cite{bochner} by the very fact that they start at degree~$\ell$
\eqref{degreel} instead of degree~0 constant term.
As with the Laguerre and Jacobi dif\/ferential equations~\eqref{Pndiffeq},
the second order dif\/ferential operator
$\widetilde{\mathcal{H}}_{\ell}(\bm{\lambda})$ is factorised into the
product of the forward shift operator $\mathcal{F}_{\ell}(\bm{\lambda})$
and the backward shift operator $\mathcal{B}_{\ell}(\bm{\lambda})$:
\begin{gather}
 \widetilde{\mathcal{H}}_{\ell}(\bm{\lambda})
  \eqdef\mathcal{B}_{\ell}(\bm{\lambda})
 \mathcal{F}_{\ell}(\bm{\lambda}),
 \label{factor}\\
 \mathcal{F}_{\ell}(\bm{\lambda}) \eqdef\cF
 \frac{\xi_{\ell}(\eta;\bm{\lambda}+\bm{\delta})}
 {\xi_{\ell}(\eta;\bm{\lambda})}
 \left(\frac{d}{d{\eta}}
 -\partial_{\eta}\log\xi_{\ell}(\eta;\bm{\lambda}+\bm{\delta})\right),
 \label{Fform}\\
 \mathcal{B}_{\ell}(\bm{\lambda}) \eqdef -4\cF^{-1}c_2(\eta)
 \frac{\xi_{\ell}(\eta;\bm{\lambda})}
 {\xi_{\ell}(\eta;\bm{\lambda}+\bm{\delta})}
 \left(\frac{d}{d{\eta}}
 +\frac{c_1(\eta,\bm{\lambda}+\ell\bm{\delta})}{c_2(\eta)}
 -\partial_{\eta}\log\xi_{\ell}(\eta;\bm{\lambda})\right).
 \label{Bform}
\end{gather}
The dif\/ferential operator $\widetilde{\mathcal{H}}_{\ell}(\bm{\lambda})$
is obtained from the factorised quantum mechanical Hamiltonian~(16) of~\cite{os16} and~(3) of~\cite{os19}, by similarity transformation.
Note that
$\frac{c_1(\eta,\bm{\lambda}+\ell\bm{\delta})}{c_2(\eta)}=
\partial_{\eta}\log{W}(\eta;\bm{\lambda}+(\ell+1)\bm{\delta})$.
It is straightforward to derive the explicit form of
$\widetilde{\mathcal{H}}_\ell(\bm{\lambda})$:
\begin{gather}
 \widetilde{\mathcal{H}}_{\ell}(\bm{\lambda})
  =-4\Biggl(c_2(\eta)\frac{d^2}{d\eta^2}
 +\bigl(c_1(\eta,\bm{\lambda}+\ell\bm{\delta})-2c_2(\eta)
 \partial_{\eta}\log\xi_{\ell}(\eta;\bm{\lambda})\bigr)\frac{d}{d\eta}\nonumber\\
  \phantom{\widetilde{\mathcal{H}}_{\ell}(\bm{\lambda})=-4\Biggl(}
 +2d_1(\bm{\lambda})\frac{c_2(\eta)}{d_2(\eta)}
 \frac{\partial_{\eta}\xi_{\ell}(\eta;\bm{\lambda}+\bm{\delta})}
 {\xi_{\ell}(\eta;\bm{\lambda})}
 +\frac14\,\widetilde{\mathcal{E}}_{\ell}(\bm{\lambda}+\bm{\delta})\Biggr).
 \label{newttildeeq}
\end{gather}
Use is made of the second order dif\/ferential equations for
$\xi_{\ell}(\eta;\bm{\lambda}+\bm{\delta})$ \eqref{xildiffeq} and
the iden\-ti\-ty~\eqref{xil(l+d)} to derive the above simple result.

For $\ell=0$ the above dif\/ferential equation \eqref{htildeeq} with
\eqref{newttildeeq} reduces to the second order dif\/ferential equation
for the Laguerre or Jacobi polynomials:
\begin{alignat}{3}
 & &&
 \widetilde{\mathcal{H}}_0(\bm{\lambda})
 =-4\left(c_2(\eta)\frac{d^2}{d\eta^2}
 +c_1(\eta,\bm{\lambda})\frac{d}{d\eta}\right),&
 \label{Ht0}\\
 & (\text{L}): \quad &&
 \widetilde{\mathcal{H}}_0(\bm{\lambda})
 =-4\left(\eta\frac{d^2}{d\eta^2}
 +(g+\tfrac12-\eta)\frac{d}{d\eta}\right), &
 \label{defHL}\\
 &&& \eta\partial_{\eta}^2P_n(\eta;g)+\big(g+\tfrac{1}{2}-\eta\big)
 \partial_{\eta}P_n(\eta;g)+n\,P_n(\eta;g)=0, &\nonumber\\
 & (\text{J}):\quad &&
 \widetilde{\mathcal{H}}_0(\bm{\lambda})
 =-4\left((1-\eta^2)\frac{d^2}{d\eta^2}
 +\bigl(h-g-(g+h+1)\eta\bigr)\frac{d}{d\eta}\right), &
 \label{defHJ}\\
&& &(1-\eta^2)\partial_{\eta}^2P_n(\eta;g,h)+\bigl(h-g-(g+h+1)\eta\bigr)
 \partial_{\eta}P_n(\eta;g,h)\nonumber &\\
 &&& \qquad {} +n(n+g+h)P_n(\eta;g,h)=0, & \nonumber
\end{alignat}
which has, as is well-known, one regular singularity at $\eta=0$ and
one irregular singularity at $\eta=\infty$ for the Laguerre case and
three regular singularities at $\eta=\pm1,\infty$ for the Jacobi case.
For a non-negative integer $\ell$, the singularity structure of the
second order dif\/ferential equation~\eqref{htildeeq} is again quite simple.
It has extra regular singularities at the $\ell$  zeros of the deforming
polyno\-mial~$\xi_{\ell}(\eta;\bm{\lambda})$:
\begin{gather*}
 \eta=\eta_j,\qquad \xi_{\ell}(\eta_j;\bm{\lambda})=0,\qquad
 j=1,2,\ldots,\ell,
\end{gather*}
and the corresponding exponents are the same for all the singular points:
\begin{gather*}
 \{\text{exponents at $\eta_j$}\}=\{0,3\},\qquad j=1,2,\ldots,\ell.
\end{gather*}
In \looseness=1 other words, the alternative solution of the second order
linear dif\/ferential equation \eqref{htildeeq} has a cubic zero at
$\eta=\eta_j$. It is singular at $\eta=0,\infty$ for the L1 and L2
and at $\eta=\pm1$ for the J1 and J2 with the exponents replaced
by $g\to g+\ell$, $h\to h+\ell$. To the best of our knowledge, the
two sets of $X_{\ell}$ orthogonal polynomials, J1 and J2 are the
f\/irst examples of global solutions of Fuchsian dif\/ferential
equations having as many as $\ell+3$ regular singularities and
forming a complete orthogonal system. The~L1 and~L2 are conf\/luent
types obtained from J1 and J2 by certain limits \eqref{J->L}. By
the way, it is elementary to show that
$\xi_{\ell}(\eta;\bm{\lambda})$ has only simple zeros.

Let us write down the explicit form of the above dif\/ferential equation
for the four cases:
\begin{alignat}{3}
 & (\text{L1}): &&
 \eta\partial_{\eta}^2P_{\ell,n}(\eta;\bm{\lambda})
 +\left(g+\ell+\tfrac12-\eta
 -2\frac{\eta\,\partial_{\eta}\xi_{\ell}(\eta;\bm{\lambda})}
 {\xi_{\ell}(\eta;\bm{\lambda})}\right)
 \partial_{\eta}P_{\ell,n}(\eta;\bm{\lambda})\nonumber & \\
 &&& \phantom{\eta\partial_{\eta}^2P_{\ell,n}(\eta;\bm{\lambda})}{}
 +\left(2\frac{\eta\,\partial_{\eta}\xi_{\ell}(\eta;\bm{\lambda}+\bm{\delta})}
 {\xi_{\ell}(\eta;\bm{\lambda})}+n-\ell\right)
 P_{\ell,n}(\eta;\bm{\lambda})=0, &
 \label{L1eq}\\
 & (\text{L2}): \quad &&
 \eta\partial_{\eta}^2P_{\ell,n}(\eta;\bm{\lambda})
 +\left(g+\ell+\tfrac12-\eta
 -2\frac{\eta\,\partial_{\eta}\xi_{\ell}(\eta;\bm{\lambda})}
 {\xi_{\ell}(\eta;\bm{\lambda})}\right)
 \partial_{\eta}P_{\ell,n}(\eta;\bm{\lambda})\nonumber & \\
 &&&
 \phantom{\eta\partial_{\eta}^2P_{\ell,n}(\eta;\bm{\lambda})}{}
 +\left(-2\frac{(g+\tfrac12)\partial_{\eta}
 \xi_{\ell}(\eta;\bm{\lambda}+\bm{\delta})}
 {\xi_{\ell}(\eta;\bm{\lambda})}+n+\ell\right)
 P_{\ell,n}(\eta;\bm{\lambda})=0,&
 \label{L2eq}\\
 & (\text{J1}):\quad &&
 \big(1-\eta^2\big)\partial_{\eta}^2P_{\ell,n}(\eta;\bm{\lambda})\nonumber &\\
 &&&
 \quad{}
 +\left(h-g-(g+h+2\ell+1)\eta
 -2\frac{(1-\eta^2)\partial_{\eta}\xi_{\ell}(\eta;\bm{\lambda})}
 {\xi_{\ell}(\eta;\bm{\lambda})}\right)
 \partial_{\eta}P_{\ell,n}(\eta;\bm{\lambda})\nonumber & \\
 &&&
 \quad{}
 +\left(-\frac{2(h+\tfrac12)(1-\eta)\partial_{\eta}
 \xi_{\ell}(\eta;\bm{\lambda}+\bm{\delta})}
 {\xi_{\ell}(\eta;\bm{\lambda})}+\ell(\ell+g-h-1)+n(n+g+h+2\ell)\right)\nonumber &\\
&&&
 \quad{}\times
 P_{\ell,n}(\eta;\bm{\lambda})=0, &
 \label{J1eq}\\
 & (\text{J2}): \quad &&
 \big(1-\eta^2\big)\partial_{\eta}^2P_{\ell,n}(\eta;\bm{\lambda})\nonumber &\\
 &&&\quad{}
 +\left(h-g-(g+h+2\ell+1)\eta-2\frac{(1-\eta^2)\partial_{\eta}
 \xi_{\ell}(\eta;\bm{\lambda})}
 {\xi_{\ell}(\eta;\bm{\lambda})}\right)
 \partial_{\eta}P_{\ell,n}(\eta;\bm{\lambda})\nonumber&\\
&& &
\quad{}
+\left(\frac{2(g+\tfrac12)(1+\eta)\partial_{\eta}
 \xi_{\ell}(\eta;\bm{\lambda}+\bm{\delta})}
 {\xi_{\ell}(\eta;\bm{\lambda})}+\ell(\ell+h-g-1)+n(n+g+h+2\ell)\right)\nonumber&\\
&& &
\quad{} \times
 P_{\ell,n}(\eta;\bm{\lambda})=0.&
 \label{J2eq}
\end{alignat}
Let us remark that the zeros of the shifted deforming polynomial
$\xi_{\ell}(\eta;\bm{\lambda}+\bm{\delta})$ are regular points.
It is straightforward to verify by direct calculation that the L1, L2,
J1 and J2 $X_{\ell}$ polynomials \eqref{XLform}--\eqref{defxiJ} for
lower $\ell$ and $n$ really satisfy the above dif\/ferential equations
\eqref{L1eq}--\eqref{J2eq}.
For analytical proof see the subsequent sections.

\section{Shape invariance}\label{section4}

This section is a reformulation of the shape invariance in the language
of ordinary dif\/ferential equations.
In one-dimensional quantum mechanics, {\em shape invariance\/} is a
suf\/f\/icient condition~\cite{genden} for {\em exact solvability\/} and
it was the guiding principle for the discovery of these $X_{\ell}$
orthogonal polynomials~\cite{os16,os17,os18,os19}.
Let us introduce another second order linear dif\/ferential operator
$\widetilde{\mathcal{H}}_{\ell}^{(1)}(\bm{\lambda})$ by
interchanging the order of the two factors
$\mathcal{F}_{\ell}(\bm{\lambda})$ and $\mathcal{B}_{\ell}(\bm{\lambda})$
of $\widetilde{\mathcal{H}}_{\ell}(\bm{\lambda})$ \eqref{factor}:
\begin{gather*}
 \widetilde{\mathcal{H}}_{\ell}^{(1)}(\bm{\lambda})\eqdef
 \mathcal{F}_{\ell}(\bm{\lambda})
 \mathcal{B}_{\ell}(\bm{\lambda}).
\end{gather*}
It is obvious that these two operators are intertwined by
$\mathcal{F}_{\ell}(\bm{\lambda})$ and $\mathcal{B}_{\ell}(\bm{\lambda})$:
\begin{gather*}
 \mathcal{F}_{\ell}(\bm{\lambda})\mathcal{B}_{\ell}(\bm{\lambda})
 \mathcal{F}_{\ell}(\bm{\lambda})
  =\mathcal{F}_{\ell}(\bm{\lambda})
 \widetilde{\mathcal{H}}_{\ell}(\bm{\lambda})
 =\widetilde{\mathcal{H}}_{\ell}^{(1)}(\bm{\lambda})
 \mathcal{F}_{\ell}(\bm{\lambda}),\\
 \mathcal{B}_{\ell}(\bm{\lambda})\mathcal{F}_{\ell}(\bm{\lambda})
 \mathcal{B}_{\ell}(\bm{\lambda})
=\widetilde{\mathcal{H}}_{\ell}(\bm{\lambda})
 \mathcal{B}_{\ell}(\bm{\lambda})
 =\mathcal{B}_{\ell}(\bm{\lambda})
 \widetilde{\mathcal{H}}_{\ell}^{(1)}(\bm{\lambda}),
\end{gather*}
which implies that these two associated linear dif\/ferential operators
$\widetilde{\mathcal{H}}_{\ell}(\bm{\lambda})$ and
$\widetilde{\mathcal{H}}_{\ell}^{(1)}(\bm{\lambda})$ are
{\em iso-spectral\/} except for the lowest eigenfunction
$P_{\ell,0}(\eta;\bm{\lambda})$ \eqref{degreel}
which is annihilated by $\mathcal{F}_{\ell}(\bm{\lambda})$, see~\eqref{Fform}:
\begin{gather*}
 \mathcal{F}_{\ell}(\bm{\lambda})\xi_{\ell}(\eta;\bm{\lambda}+\bm{\delta})
 =0\qquad
 \bigl(\Rightarrow \
 \widetilde{\mathcal{H}}_{\ell}(\bm{\lambda})
 P_{\ell,0}(\eta;\bm{\lambda})=0\bigr).
\end{gather*}
If we denote the set of eigenfunctions of
$\widetilde{\mathcal{H}}_{\ell}^{(1)}(\bm{\lambda})$ as
$\{P_{\ell,n}^{(1)}(\eta;\bm{\lambda})\}$ ($n=0,1,\ldots$), with
arbitrary normalisation, we obtain one to one correspondence of
$\{P_{\ell,n}(\eta;\bm{\lambda})\}$ and
$\{P_{\ell,n}^{(1)}(\eta;\bm{\lambda})\}$ except for the lowest
eigenfunction of $\widetilde{\mathcal{H}}_{\ell}(\bm{\lambda})$:
\begin{gather}
 \widetilde{\mathcal{H}}_{\ell}(\bm{\lambda})
 P_{\ell,n}(\eta;\bm{\lambda})
  =\mathcal{E}_{\ell,n}(\bm{\lambda})P_{\ell,n}(\eta;\bm{\lambda}),
 \qquad  n=0,1,\ldots,\nonumber\\
 \widetilde{\mathcal{H}}_{\ell}^{(1)}(\bm{\lambda})
 P_{\ell,n-1}^{(1)}(\eta;\bm{\lambda})
 =\mathcal{E}_{\ell,n}(\bm{\lambda})P_{\ell,n-1}^{(1)}(\eta;\bm{\lambda}),
 \qquad n=1,2,\ldots,\nonumber\\
 \mathcal{F}_{\ell}(\bm{\lambda})P_{\ell,n}(\eta;\bm{\lambda})
 \propto P_{\ell,n-1}^{(1)}(\eta;\bm{\lambda}),\qquad
 \mathcal{B}_{\ell}(\bm{\lambda})P_{\ell,n-1}^{(1)}(\eta;\bm{\lambda})
 \propto P_{\ell,n}(\eta;\bm{\lambda}).
 \label{fbrelation}
\end{gather}
This much is a trivial consequence of the factorisation of
$\widetilde{\mathcal{H}}_{\ell}(\bm{\lambda})$ \eqref{factor}.
The essential property of the two associated dif\/ferential operators
$\widetilde{\mathcal{H}}_{\ell}(\bm{\lambda})$ and
$\widetilde{\mathcal{H}}_{\ell}^{(1)}(\bm{\lambda})$ is that
$\widetilde{\mathcal{H}}_{\ell}^{(1)}(\bm{\lambda})$
{\em has the same shape as\/}
$\widetilde{\mathcal{H}}_{\ell}(\bm{\lambda})$ with shifted
parameters, $\bm{\lambda}\to\bm{\lambda}+\bm{\delta}$, and an additive
constant corresponding to the lowest eigenvalue measured from the bottom:
\begin{gather}
 \widetilde{\mathcal{H}}_{\ell}^{(1)}(\bm{\lambda})
  =\widetilde{\mathcal{H}}_{\ell}(\bm{\lambda}+\bm{\delta})
 +\mathcal{E}_{1}(\bm{\lambda}+\ell\bm{\delta}), \qquad \text{or} \nonumber\\
 \mathcal{F}_{\ell}(\bm{\lambda})\mathcal{B}_{\ell}(\bm{\lambda})
  =\mathcal{B}_{\ell}(\bm{\lambda}+\bm{\delta})
 \mathcal{F}_{\ell}(\bm{\lambda}+\bm{\delta})
 +\mathcal{E}_{1}(\bm{\lambda}+\ell\bm{\delta}).
 \label{defshapeinv}
\end{gather}
The system is called shape invariant if the above condition is satisf\/ied.
See \cite{os16,os19, os18} for the def\/inition of shape invariance within
the framework of quantum mechanics. The above is an equivalent def\/inition
within the framework of the Sturm--Liouville theory.

Shape invariance allows us to choose the normalisation of
$\{P_{\ell,n}^{(1)}(\eta;\bm{\lambda})\}$ to achieve
\begin{gather*}
 P_{\ell,n}^{(1)}(\eta;\bm{\lambda})
 =P_{\ell,n}(\eta;\bm{\lambda}+\bm{\delta}),\qquad n=0,1,\ldots .
\end{gather*}
Then the above two relations \eqref{fbrelation}, the
{\em forward and backward shift relations\/}, give the constraints
on the functional forms of $\{P_{\ell,n}(\eta;\bm{\lambda})\}$:
\begin{gather}
\mathcal{F}_{\ell}(\bm{\lambda})P_{\ell,n}(\eta;\bm{\lambda})
 =f_n(\bm{\lambda}+\ell\bm{\delta})
 P_{\ell,n-1}(\eta;\bm{\lambda}+\bm{\delta}),\qquad n=0,1,\ldots,
 \label{FonP}\\
\mathcal{B}_{\ell}(\bm{\lambda})
 P_{\ell,n-1}(\eta;\bm{\lambda}+\bm{\delta})
 =b_{n-1}(\bm{\lambda}+\ell\bm{\delta})P_{\ell,n}(\eta;\bm{\lambda}),
 \qquad n=1,2,\ldots,
 \label{BonP}
\end{gather}
where $f_n(\bm{\lambda})$ and $b_{n-1}(\bm{\lambda})$ are given in~\eqref{CFfnbn-1}.
These amount to a version of {\em Rodrigues formula\/} expressing
$P_{\ell,n}(\eta;\bm{\lambda})$ in terms of repeated application of
the backward shift operators on the lowest degree eigenpolynomials
with the $n$-th shifted parameters,
$P_{\ell,0}(\eta;\bm{\lambda}+n\bm{\delta})
=\xi_{\ell}(\eta;\bm{\lambda}+(n+1)\bm{\delta})$:
\begin{gather*}
 P_{\ell,n}(\eta;\bm{\lambda})
 =\prod_{k=0}^{n-1}\frac{\mathcal{B}_{\ell}(\bm{\lambda}+k\bm{\delta})}
 {b_{n-k-1}(\bm{\lambda}+(\ell+k)\bm{\delta})}
 \cdot\xi_{\ell}(\eta;\bm{\lambda}+(n+1)\bm{\delta}),
\end{gather*}
where $\prod\limits_{k=0}^{n-1}a_k=a_0a_1\cdots a_{n-1}$.
For the $X_{\ell}$ Laguerre and Jacobi polynomials, the Rodrigues
formula reads explicitly:
\begin{alignat*}{3}
& (\text{L}):\quad && P_{\ell,n}(\eta;\bm{\lambda})
  =\frac{1}{n!}
 \frac{\xi_{\ell}(\eta;\bm{\lambda})}{e^{-\eta}\eta^{g+\ell-\frac12}}
 \prod_{k=0}^{n-1}\left(\frac{d}{d\eta}+\partial_{\eta}\log
 \frac{\xi_{\ell}(\eta;\bm{\lambda}+(k+1)\bm{\delta})}
 {\xi_{\ell}(\eta;\bm{\lambda}+k\bm{\delta})}\right) & \nonumber\\
 &&& \phantom{P_{\ell,n}(\eta;\bm{\lambda})=}{} \times
 \frac{e^{-\eta}\eta^{n+g+\ell-\frac12}}
 {\xi_{\ell}(\eta;\bm{\lambda}+n\bm{\delta})}
 \xi_{\ell}\bigl(\eta;\bm{\lambda}+(n+1)\bm{\delta}\bigr),\\
&(\text{J}):\quad && P_{\ell,n}(\eta;\bm{\lambda})
  =\frac{(-1)^n}{2^nn!}
 \frac{\xi_{\ell}(\eta;\bm{\lambda})}
 {(1-\eta)^{g+\ell-\frac12}(1+\eta)^{h+\ell-\frac12}}
 \prod_{k=0}^{n-1}\!\left(\frac{d}{d\eta}+\partial_{\eta}\log
 \frac{\xi_{\ell}(\eta;\bm{\lambda}+(k+1)\bm{\delta})}
 {\xi_{\ell}(\eta;\bm{\lambda}+k\bm{\delta})}\right)& \nonumber\\
&&& \phantom{P_{\ell,n}(\eta;\bm{\lambda})=}{} \times
 \frac{(1-\eta)^{n+g+\ell-\frac12}(1+\eta)^{n+h+\ell-\frac12}}
 {\xi_{\ell}(\eta;\bm{\lambda}+n\bm{\delta})}
 \xi_{\ell}\bigl(\eta;\bm{\lambda}+(n+1)\bm{\delta}\bigr). &
\end{alignat*}

For $\ell=0$, $\xi_0(\eta;\bm{\lambda})=1$, the above two formulas
reduce to the well-known Rodrigues formulas for the Laguerre and Jacobi
polynomials,~\eqref{LRod} and~\eqref{JRod}.
For $\ell=1$ these two formulas are equivalent to the Rodrigues-type
formulas~(77) and~(52) in Gomez-Ullate et al.~\cite{gomez1}.
For lower $\ell$, it is straightforward to verify the shape invariance
relation \eqref{defshapeinv} by direct calculation.
In~\cite{os18} it was shown that the shape invariance relation is
attributed to a new polynomial identity of degree $3\ell$ involving
cubic products of the Laguerre or Jacobi polynomials. These identities
are proved elementarily by combining simple identities in~\cite{os18}.

It is a good challenge to derive more explicit expressions of the Rodrigues
formulas for
each of the four families of the exceptional orthogonal polynomials.

\section{Forward and backward shift relations}\label{section5}

Again it is straightforward to verify that the above Rodrigues formulas
provide the explicit forms of the four families of the $X_{\ell}$
polynomials \eqref{XLform}--\eqref{defxiJ} for lower $\ell$ and $n$
by direct calculation.
For analytic proof, one needs to verify that the above forward
\eqref{FonP} and backward \eqref{BonP} shift relations are actually
satisf\/ied by the four families of the $X_{\ell}$ polynomials
\eqref{XLform}--\eqref{defxiJ}.
The forward and backward shift relations are attributed to new polynomial
identities of degree $2\ell+n-1$ and $2\ell+n$ involving cubic products
of the Laguerre or Jacobi polynomials in a similar way to the shape
invariance relations \cite{os18}.
The forward (F) \eqref{FonP} and backward (B) \eqref{BonP} shift relations
read explicitly
\begin{alignat}{3}
&  (\text{F}):\quad &&
 0 =\cF\left(\xi_{\ell}(\eta;\bm{\lambda}+\bm{\delta})\frac{d}{d\eta}
 -\partial_{\eta}\xi_{\ell}(\eta;\bm{\lambda}+\bm{\delta})\right)
 P_{\ell,n}(\eta;\bm{\lambda})\nonumber & \\
&&& \phantom{0}{}
 -f_n(\bm{\lambda}+\ell\bm{\delta})\xi_{\ell}(\eta;\bm{\lambda})
 P_{\ell,n-1}(\eta;\bm{\lambda}+\bm{\delta}), &
 \label{for_rel}\\
& (\text{B}):\quad &&
 0 =\left(\xi_{\ell}(\eta;\bm{\lambda})\bigl(c_2(\eta)\frac{d}{d\eta}
 +c_1(\eta,\bm{\lambda}+\ell\bm{\delta})\bigr)
 -c_2(\eta)\partial_{\eta}\xi_{\ell}(\eta;\bm{\lambda})\right)
 P_{\ell,n-1}(\eta;\bm{\lambda}+\bm{\delta})\nonumber & \\
&& &\phantom{0=}{}
 +\tfrac14\cF b_{n-1}(\bm{\lambda}+\ell\bm{\delta})
 \xi_{\ell}(\eta;\bm{\lambda}+\bm{\delta})
 P_{\ell,n}(\eta;\bm{\lambda}), &
 \label{back_rel}
\end{alignat}
where $P_{\ell,n}$ is given in \eqref{Plnform}.

The new polynomial identities are as follows:

\noindent
({\bf L1}): ($\alpha=g+\ell-\frac12$)
\begin{alignat}{3}
& (\text{F}):\quad &&
 0 =\bigl(L_{\ell}^{(\alpha)}(-x)\tfrac{d}{dx}
 -\partial_xL_{\ell}^{(\alpha)}(-x)\bigr)
 \bigl(L_{\ell}^{(\alpha)}(-x)L_n^{(\alpha-1)}(x)
 -L_{\ell}^{(\alpha-1)}(-x)\partial_xL_n^{(\alpha-1)}(x)\bigr) & \nonumber\\
&&& \phantom{0=}{}
 +L_{\ell}^{(\alpha-1)}(-x)
 \bigl(L_{\ell}^{(\alpha+1)}(-x)L_{n-1}^{(\alpha)}(x)
 -L_{\ell}^{(\alpha)}(-x)\partial_xL_{n-1}^{(\alpha)}(x)\bigr), &
 \label{fidL1}\\
& (\text{B}):\quad &&
 0 =\bigl(L_{\ell}^{(\alpha-1)}(-x)(x\tfrac{d}{dx}+\alpha+1-x)
 -x\partial_xL_{\ell}^{(\alpha-1)}(-x)\bigr) & \nonumber\\
&&& \phantom{0=}{} \times
 \bigl(L_{\ell}^{(\alpha+1)}(-x)L_{n-1}^{(\alpha)}(x)
 -L_{\ell}^{(\alpha)}(-x)\partial_xL_{n-1}^{(\alpha)}(x)\bigr) & \nonumber\\
&&& \phantom{0=}{}
 -nL_{\ell}^{(\alpha)}(-x)
 \bigl(L_{\ell}^{(\alpha)}(-x)L_n^{(\alpha-1)}(x)
 -L_{\ell}^{(\alpha-1)}(-x)\partial_xL_n^{(\alpha-1)}(x)\bigr), &
 \label{bidL1}
\end{alignat}
({\bf L2}): ($\alpha=g+\ell-\frac12$)
\begin{alignat}{3}
 & (\text{F}):\quad &&
 0 =\bigl(L_{\ell}^{(-\alpha-2)}(x)\tfrac{d}{dx}
 -\partial_xL_{\ell}^{(-\alpha-2)}(x)\bigr) &\nonumber\\
&&& \phantom{0=}{}\times
 \bigl((\alpha-\ell+1)L_{\ell}^{(-\alpha-2)}(x)L_n^{(\alpha+1)}(x)
 +xL_{\ell}^{(-\alpha-1)}(x)\partial_xL_n^{(\alpha+1)}(x)\bigr) & \label{fidL2}\\
&&& \phantom{0=}{} \quad
 +L_{\ell}^{(-\alpha-1)}(x)
 \bigl((\alpha-\ell+2)L_{\ell}^{(-\alpha-3)}(x)L_{n-1}^{(\alpha+2)}(x)
 +xL_{\ell}^{(-\alpha-2)}(x)\partial_xL_{n-1}^{(\alpha+2)}(x)\bigr), & \nonumber
\\
& (\text{B}):\quad &&
 0 =\bigl(L_{\ell}^{(-\alpha-1)}(x)(x\tfrac{d}{dx}+\alpha+1-x)
 -x\partial_xL_{\ell}^{(-\alpha-1)}(x)\bigr) & \nonumber\\
&&& \phantom{0=}{}\times
 \bigl((\alpha-\ell+2)L_{\ell}^{(-\alpha-3)}(x)L_{n-1}^{(\alpha+2)}(x)
 +xL_{\ell}^{(-\alpha-2)}(x)\partial_xL_{n-1}^{(\alpha+2)}(x)\bigr) & \label{bidL2}\\
&&&\phantom{0=}{}
 -nL_{\ell}^{(-\alpha-2)}(x)
 \bigl((\alpha-\ell+1)L_{\ell}^{(-\alpha-2)}(x)L_n^{(\alpha+1)}(x)
 +xL_{\ell}^{(-\alpha-1)}(x)\partial_xL_n^{(\alpha+1)}(x)\bigr),& \nonumber
\end{alignat}
({\bf J2}): ($\alpha=g+\ell-\frac12$, $\beta=h+\ell-\frac12$)
\begin{alignat}{3}
 & (\text{F}): \quad &&
  0 =\bigl(P_{\ell}^{(-\alpha-2,\beta)}(x)\tfrac{d}{dx}
 -\partial_xP_{\ell}^{(-\alpha-2,\beta)}(x)\bigr)
 \bigl((\alpha-\ell+1)P_{\ell}^{(-\alpha-2,\beta)}(x)
 P_n^{(\alpha+1,\beta-1)}(x)&\nonumber\\
&&&\phantom{0=}{} -(1-x)P_{\ell}^{(-\alpha-1,\beta-1)}(x)
 \partial_xP_n^{(\alpha+1,\beta-1)}(x)\bigr)
 -\tfrac12(n+\alpha+\beta+1)P_{\ell}^{(-\alpha-1,\beta-1)}(x) &
\nonumber\\
&&&\phantom{0=}{}
\times
 \bigl((\alpha-\ell+2)P_{\ell}^{(-\alpha-3,\beta+1)}(x)
 P_{n-1}^{(\alpha+2,\beta)}(x)&\nonumber\\
&&&\phantom{0=}{}
 -(1-x)P_{\ell}^{(-\alpha-2,\beta)}(x)
 \partial_xP_{n-1}^{(\alpha+2,\beta)}(x)\bigr), &  \label{fidJ}\\
& (\text{B}): \quad &&
  0 = \Bigl( P_{\ell}^{(-\alpha-1,\beta-1)}(x)
 \bigl((1-x^2)\tfrac{d}{dx}+\beta-\alpha- (\alpha+\beta+2)x\bigr) & \nonumber\\
&&& \phantom{0 =}{}
 -\big(1-x^2\big)\partial_xP_{\ell}^{(-\alpha-1,\beta-1)}(x) \Bigr)
 \bigl((\alpha-\ell+2)P_{\ell}^{(-\alpha-3,\beta+1)}(x)
 P_{n-1}^{(\alpha+2,\beta)}(x)& \nonumber\\
&&& \phantom{0 =}{}
 -(1-x)P_{\ell}^{(-\alpha-2,\beta)}(x)
 \partial_xP_{n-1}^{(\alpha+2,\beta)}(x)\bigr)
 +2nP_{\ell}^{(-\alpha-2,\beta)}(x) &
\nonumber\\
&&& \phantom{0 =}{}
 \times
 \bigl((\alpha-\ell+1)P_{\ell}^{(-\alpha-2,\beta)}(x)
 P_n^{(\alpha+1,\beta-1)}(x)\nonumber\\
 &&& \phantom{0 =}{}
 -(1-x)P_{\ell}^{(-\alpha-1,\beta-1)}(x)
 \partial_xP_n^{(\alpha+1,\beta-1)}(x)\bigr). &
 \label{bidJ}
\end{alignat}
We do not show those for the J1 polynomials, since J1 and J2 are
related by the mirror image, $\eta\leftrightarrow -\eta$ and renaming
of the coupling constants $g\leftrightarrow h$.
Thus the corresponding identities are essentially the same.
The derivative terms in the above identities \eqref{fidL1}--\eqref{bidJ}
can be eliminated by using the forward shift relations \eqref{Lforward}
and \eqref{Jforward} for the Laguerre and Jacobi polynomials.

In Appendix \ref{appendixB}, the forward and backward shift relations
for the four types of $X_{\ell}$ polynomials are proven elementarily.

\section{Invariant polynomial subspace}\label{section6}

Another characterisation of polynomials satisfying dif\/ferential equations
is the existence of {\em invariant polynomial subspaces\/} \cite{gomez2}.
For the Laguerre and Jacobi dif\/ferential equations with
$\widetilde{\mathcal{H}}_0(\bm{\lambda})$ \eqref{defHL},
\eqref{defHJ}, the space $\mathcal{V}_{0,n}$ of degree $n$ polynomials
in $\eta$ is invariant
\begin{gather}
  \widetilde{\mathcal{H}}_0(\bm{\lambda})\mathcal{V}_{0,n}
 \subseteq\mathcal{V}_{0,n},\qquad
 \mathcal{V}_{0,n}\eqdef\text{Span}[1,\eta,\ldots,\eta^n],\nonumber\\
\widetilde{\mathcal{H}}_0(\bm{\lambda})\eta^n
 =\mathcal{E}_n(\bm{\lambda})\eta^n+\text{lower orders}.
 \label{hamzeroact}
\end{gather}
However, $\mathcal{V}_{0,n}$ is obviously not invariant under
$\widetilde{\mathcal{H}}_{\ell}(\bm{\lambda})$ \eqref{htildeeq},
\eqref{L1eq}--\eqref{J2eq}:
\begin{gather}
 \widetilde{\mathcal{H}}_{\ell}(\bm{\lambda})
 \mathcal{V}_{0,n}\not\subseteq \mathcal{V}_{0,n},\qquad
 \ell=1,2,\ldots.
 \label{vnotinv}
\end{gather}
Instead we have
\begin{gather}
 \widetilde{\mathcal{H}}_{\ell}(\bm{\lambda})
 \mathcal{V}_{\ell,n}\subseteq \mathcal{V}_{\ell,n},\qquad
 \ell=1,2,\ldots,\nonumber\\
\mathcal{V}_{\ell,n}\eqdef
 \begin{cases}
 \text{Span}\bigl[\eta^k\xi_{\ell}(\eta;g+1)
 -k\eta^{k-1}\xi_{\ell}(\eta;g);k=0,1,\ldots,n\bigr],\quad & (\text{L1})\vspace{1mm}\\
 \text{Span}\bigl[\eta^k\bigl((g+\tfrac12)\xi_{\ell}(\eta;g+1)
 +k\xi_{\ell}(\eta;g)\bigr);k=0,1,\ldots,n\bigr],\quad & (\text{L2})
 \end{cases}
 \label{Linv}\\
\mathcal{V}_{\ell,n}\eqdef
 \begin{cases}
 \text{Span}\bigl[\eta^k(h+\tfrac12)\xi_{\ell}(\eta;g+1,h+1) & \\
 \hspace*{25mm}{}
 +k(1+\eta)\eta^{k-1}\xi_{\ell}(\eta;g,h);k=0,1,\ldots,n\bigr],\quad  & (\text{J1})\\
 \text{Span}\bigl[\eta^k(g+\tfrac12)\xi_{\ell}(\eta;g+1,h+1)& \\
 \hspace*{25mm}{}
 -k(1-\eta)\eta^{k-1}\xi_{\ell}(\eta;g,h);k=0,1,\ldots,n\bigr].\quad & (\text{J2})
 \end{cases}
 \label{Jinv}
\end{gather}
As the basis vectors of the invariant polynomial subspace
$\mathcal{V}_{\ell,n}$ for the $X_{\ell}$ Jacobi polynomials, one could
have chosen for $k=0,1,\ldots,n$,
\begin{alignat}{3}
& (1-\eta)^{k-1}\bigl((h+\tfrac12)(1-\eta)\xi_{\ell}(\eta;g+1,h+1)
 -k(1+\eta)\xi_{\ell}(\eta;g,h)\bigr), \quad &&  (\text{J1}) &
 \label{J1inv3}\\
 &(1+\eta)^k\bigl((h+\tfrac12)\xi_{\ell}(\eta;g+1,h+1)
 +k\,\xi_{\ell}(\eta;g,h)\bigr),\quad&& (\text{J1}) &
 \label{J1inv2}\\
 &(1-\eta)^k\bigl((g+\tfrac12)\xi_{\ell}(\eta;g+1,h+1)
 +k\,\xi_{\ell}(\eta;g,h)\bigr).\quad&& (\text{J2})&
 \label{J2inv2}
\end{alignat}
After overall rescaling, \eqref{J1inv3} and \eqref{J2inv2} go to
those of the $X_{\ell}$ Laguerre polynomials \eqref{Linv} in the
limit \eqref{J->L}, as J1$\to$L1 and J2$\to$L2. As will be shown
shortly these basis vectors \eqref{J1inv2} and \eqref{J2inv2} have
simpler integration formulas than \eqref{Jinv}. It should be
stressed that these basis vectors have the same common structure
as the $X_{\ell}$ polynomials \eqref{Pln=Xil[Pn]}, \eqref{Xil},
\begin{gather}
 \Xil{p_k(\eta)}
 =d_1(\bm{\lambda})\xi_{\ell}(\eta;\bm{\lambda}+\bm{\delta})p_k(\eta)\!
 -d_2(\eta)\xi_{\ell}(\eta;\bm{\lambda})
 \partial_{\eta}p_k(\eta)
 \in\mathcal{V}_{\ell,n},\!\!\qquad k=0,1,\ldots,n,\!\!\!
 \label{genbasis}
\end{gather}
in which $p_k(\eta)$ is an arbitrary degree $k$ polynomial in $\eta$.
These basis vectors are so chosen as not to develop any singularities at
the zeros of $\xi_{\ell}(\eta;\bm{\lambda})$ when applied by the forward
shift operator $\mathcal{F}_{\ell}(\bm{\lambda})$ \eqref{Fform}.
In fact, for a polynomial $p(\eta)$, $\mathcal{F}_{\ell}(\bm{\lambda})$
acts on $\Xil{p(\eta)}$ as
\begin{gather}
  \cF^{-1}\mathcal{F}_{\ell}(\bm{\lambda})\Xil{p(\eta)}=
\bigl(d_1(\bm{\lambda}+\bm{\delta})\partial_{\eta}p(\eta)
 -\partial_{\eta}\bigl(d_2(\eta)\partial_{\eta}p(\eta)\bigr)\bigr)
 \xi_{\ell}(\eta;\bm{\lambda}+\bm{\delta})\nonumber\\
 \phantom{\cF^{-1}\mathcal{F}_{\ell}(\bm{\lambda})\Xil{p(\eta)}= }{}
 +d_2(\eta)\partial_{\eta}p(\eta)
 \partial_{\eta}\xi_{\ell}(\eta;\bm{\lambda}+\bm{\delta}),
 \label{FXil[p]}
\end{gather}
where we have used \eqref{xil(l+d)} to eliminate
$\partial_{\eta}\xi_{\ell}(\eta;\bm{\lambda})$.
Since the backward shift operator
$\mathcal{B}_{\ell}(\bm{\lambda})$ \eqref{Bform} does not cause any
singularity at the zeros of $\xi_{\ell}(\eta;\bm{\lambda})$ and
the operator $\widetilde{\mathcal{H}}_{\ell}(\bm{\lambda})
=\mathcal{B}_{\ell}(\bm{\lambda})\mathcal{F}_{\ell}(\bm{\lambda})$
has the form \eqref{newttildeeq}, the application of
$\widetilde{\mathcal{H}}_{\ell}(\bm{\lambda})$ on these basis vectors
will result in polynomials in $\eta$.

Let us evaluate the action of the second order dif\/ferential operator
$\widetilde{\mathcal{H}}_{\ell}(\bm{\lambda})$ on $\Xil{p(\eta)}$ by
applying the backward shift operator $\mathcal{B}_\ell(\bm{\lambda})$
on \eqref{FXil[p]}:
\begin{gather*}
 \widetilde{\mathcal{H}}_{\ell}(\bm{\lambda})\Xil{p(\eta)}
 =\xi_{\ell}(\eta;\bm{\lambda})X(\eta)
 +\xi_{\ell}(\eta;\bm{\lambda}+\bm{\delta})Y(\eta).
\end{gather*}
The coef\/f\/icients $X(\eta)$ and $Y(\eta)$ can be written as
\begin{gather*}
 X(\eta) =4d_2(\eta)\partial_{\eta}\bigl(
 c_2(\eta)\partial_{\eta}^2p(\eta)+c_1(\eta,\bm{\lambda'})
 \partial_{\eta}p(\eta)\bigr)
 =-d_2(\eta)\partial_{\eta}\bigl(
 \widetilde{\mathcal{H}}_0(\bm{\lambda'})p(\eta)\bigr),\\
 Y(\eta) =-4d_1(\bm{\lambda})\bigl(
 c_2(\eta)\partial_{\eta}^2p(\eta)+c_1(\eta,\bm{\lambda'})
 \partial_{\eta}p(\eta)\bigr)
 =d_1(\bm{\lambda})\widetilde{\mathcal{H}}_0(\bm{\lambda'})p(\eta),
\end{gather*}
where $\bm{\lambda'}=\bm{\lambda}+\ell\bm{\delta}+\bm{\tilde{\delta}}$
and $\widetilde{\mathcal{H}}_0(\bm{\lambda})$ is def\/ined in~\eqref{Ht0}.
Therefore the results can be expressed in quite a simple form as
\begin{gather*}
 \widetilde{\mathcal{H}}_{\ell}(\bm{\lambda})\Xil{p(\eta)}
 =\Xilbig{\widetilde{\mathcal{H}}_0(\bm{\lambda}+\ell\bm{\delta}
 +\bm{\tilde{\delta}})p(\eta)}.
\end{gather*}
By taking $p(\eta)=p_k(\eta)$, the above basis vectors satisfy
\begin{gather*}
 \widetilde{\mathcal{H}}_{\ell}(\bm{\lambda})\Xil{p_k(\eta)}
 =\Xilbig{\widetilde{\mathcal{H}}_0(\bm{\lambda}+\ell\bm{\delta}
 +\bm{\tilde{\delta}})p_k(\eta)}
 \in\mathcal{V}_{\ell,n},\qquad k\le n.
\end{gather*}
In other words, we have shown that $\mathcal{V}_{\ell,n}$ is an
invariant polynomial subspace of the dif\/ferential operator
$\widetilde{\mathcal{H}}_{\ell}(\bm{\lambda})$ and we obtain,
corresponding to \eqref{hamzeroact},
\begin{gather*}
 \widetilde{\mathcal{H}}_{\ell}(\bm{\lambda})\Xil{p_n(\eta)}=
 \mathcal{E}_n(\bm{\lambda}+\ell\bm{\delta})\Xil{p_n(\eta)}
 +\text{lower orders}.
\end{gather*}
In particular, if we choose $p_n(\eta)$ as the eigenfunction
$P_n(\eta;\bm{\lambda}+\ell\bm{\delta}+\tilde{\bm{\delta}})$ of
$\widetilde{\mathcal{H}}_0(\bm{\lambda}+\ell\bm{\delta}
+\tilde{\bm{\delta}})$,
\begin{gather*}
 \widetilde{\mathcal{H}}_0(\bm{\lambda}+\ell\bm{\delta}
 +\tilde{\bm{\delta}})
 P_n(\eta;\bm{\lambda}+\ell\bm{\delta}+\tilde{\bm{\delta}})
 =\mathcal{E}_n(\bm{\lambda}+\ell\bm{\delta})
 P_n(\eta;\bm{\lambda}+\ell\bm{\delta}+\tilde{\bm{\delta}}),
\end{gather*}
then we f\/ind that
$\Xil{P_n(\eta;\bm{\lambda}+\ell\bm{\delta}+\tilde{\bm{\delta}})}$
is the eigenfunction of $\widetilde{\mathcal{H}}_{\ell}(\bm{\lambda})$,
\begin{gather*}
 \widetilde{\mathcal{H}}_{\ell}(\bm{\lambda})
 \Xilbig{P_n(\eta;\bm{\lambda}+\ell\bm{\delta}+\tilde{\bm{\delta}})}
 =\mathcal{E}_n(\bm{\lambda}+\ell\bm{\delta})
 \Xilbig{P_n(\eta;\bm{\lambda}+\ell\bm{\delta}+\tilde{\bm{\delta}})},
\end{gather*}
as given in \eqref{Pln=Xil[Pn]}.
Note that the ef\/fect of $\tilde{\bm{\delta}}$ in the eigenvalue cancels
out. This is another analytical proof for the explicit forms of the
$X_{\ell}$ polynomials $P_{\ell,n}(\eta;\bm{\lambda})$ as given
\eqref{Pln=Xil[Pn]}.

\section{Integration formulas}\label{section7}

Another well-known construction method of orthogonal polynomials is
the Gram--Schmidt orthonormalisation of certain basis vectors under
a given inner product specif\/ied by a weight function.
Let us introduce two types of inner products
$\langle *,*\rangle_{\ell,\bm{\lambda}}$ and $(*,*)_{\bm{\lambda}}$:
\begin{gather*}
 \bigl\langle p(\eta),q(\eta)\bigr\rangle_{\ell,\bm{\lambda}}\eqdef
 \int p(\eta)q(\eta)\mathcal{W}_{\ell}(\eta;\bm{\lambda})d\eta,
 \qquad
 \bigl(p(\eta),q(\eta)\bigr)_{\bm{\lambda}}\eqdef
 \int p(\eta)q(\eta)W(\eta;\bm{\lambda})d\eta,
\end{gather*}
in which $p(\eta)$ and $q(\eta)$ are arbitrary functions and
$\mathcal{W}_{\ell}(\eta;\bm{\lambda})$ is the weight function for
the $X_{\ell}$ polynomials, whereas $W(\eta;\bm{\lambda})$ is the
weight function for the Laguerre or Jacobi polyno\-mials~\eqref{defweight}.

Here we present the integration formulas:
\begin{gather}
 \bigl\langle\Xil{p(\eta)},\Xil{q(\eta)}\bigr\rangle_{\ell,\bm{\lambda}}\nonumber\\
 \qquad{}
 =d_1(\bm{\lambda})d_3(\bm{\lambda}+\ell\bm{\delta},\ell)
 \bigl(p(\eta),q(\eta)\bigr)_{\bm{\lambda'}}
 +\tfrac14\cF^2\bigl(\partial_{\eta}p(\eta),
 \partial_{\eta}q(\eta)\bigr)_{\bm{\lambda'}+\bm{\delta}}  , \qquad \text{or}
 \label{innerform}\\
 \int\Xil{p(\eta)}\,\Xil{q(\eta)}
 \mathcal{W}_{\ell}(\eta;\bm{\lambda})d\eta\nonumber\\
\qquad {}=\int
 \Bigl(d_1(\bm{\lambda})d_3(\bm{\lambda}+\ell\bm{\delta},\ell)
 p(\eta)q(\eta)W(\eta;\bm{\lambda'})
 +\tfrac14\cF^2\,\partial_{\eta}p(\eta)\partial_{\eta}q(\eta)
 W(\eta;\bm{\lambda'}+\bm{\delta})\Bigr)d\eta,
 \label{intformula}
\end{gather}
where $\bm{\lambda'}=\bm{\lambda}+\ell\bm{\delta}+\bm{\tilde{\delta}}$.
For a proof, see Appendix~\ref{appendixC}.

With this formula one can easily verify the orthogonality relation
and the normalisation constants of the $X_{\ell}$ polynomials
\eqref{orthonormalform}, \eqref{normL}, \eqref{normJ} given in
Section~\ref{section2}. One simply takes
$p(\eta)=P_n(\eta;\bm{\lambda'})$ and
$q(\eta)=P_m(\eta;\bm{\lambda'})$ for L1--J2. Then the two terms
in \eqref{intformula} read
\begin{gather*}
 \bigl(P_n(\eta;\bm{\lambda'}),
 P_m(\eta;\bm{\lambda'})\bigr)_{\bm{\lambda'}}
 =\int P_n(\eta;\bm{\lambda'})P_m(\eta;\bm{\lambda'})W(\eta;\bm{\lambda'})
 d\eta=h_{n}(\bm{\lambda'})\delta_{nm},\\
 \cF^2\int\partial_{\eta}P_n(\eta;\bm{\lambda'})
 \partial_{\eta}P_m(\eta;\bm{\lambda'})W(\eta;\bm{\lambda'}
 +\bm{\delta})d\eta
 =f_n(\bm{\lambda'})^2
 h_{n-1}(\bm{\lambda'}+\bm{{\delta}})\delta_{nm},
\end{gather*}
where we have used \eqref{Pnforward}.
Therefore, from \eqref{Pln=Xil[Pn]}, the normalisation constant $h_{\ell,n}$
is expressed as
\begin{gather*}
 d_0(n,\bm{\lambda})^2h_{\ell,n}(\bm{\lambda})
 =d_1(\bm{\lambda})d_3(\bm{\lambda}+\ell\bm{\delta},\ell)
 h_n(\bm{\lambda'})
 +\tfrac14f_n(\bm{\lambda'})^2h_{n-1}(\bm{\lambda'}+\bm{\delta}),
\end{gather*}
where $\bm{\lambda'}=\bm{\lambda}+\ell\bm{\delta}+\bm{\tilde{\delta}}$
and $h_{-1}(\bm{\lambda})\eqdef 0$.
This formula with $h_n$ \eqref{normL0} and \eqref{normJ0} gives~\eqref{normL} and~\eqref{normJ}.

The integration formula reveals another important property of the
invariant subspaces, the orthogonality,
$\mathcal{V}_{\ell,m}\perp P_{\ell,n}(\eta;\bm{\lambda})$, $m<n$:
\begin{gather*}
 \bigl\langle\Xil{p_m(\eta)},P_{\ell,n}(\eta;\bm{\lambda})
 \bigr\rangle_{\ell,\bm{\lambda}}=0,\qquad m<n,
\end{gather*}
which is a simple consequence of \eqref{Pln=Xil[Pn]} and the well known fact
\begin{gather*}
 \bigl(p_m(\eta),P_n(\eta;\bm{\lambda})\bigr)_{\bm{\lambda}}=0,\qquad m<n.
\end{gather*}

\section[Gram-Schmidt orthonormalisation]{Gram--Schmidt orthonormalisation}\label{section8}

{\sloppy As for the direct application of Gram--Schmidt orthonormalisation, one
orthonormalises $\{\Xil{p_n(\eta)}\}$ in \eqref{genbasis} with respect
to the inner product $\langle *,*\rangle_{\ell,\bm{\lambda}}$.
The following choice of the function $p_n$ is made:
\begin{gather}
 p_n(\eta)=
 \begin{cases}
 \eta^n,\quad & (\text{L})\\
 (1\pm\eta)^n, \quad & (\text{J1/J2})
 \end{cases}
 \label{GSpn}
\end{gather}
with  which
 the inner products $(p_n(\eta), p_m(\eta))_{\bm{\lambda'}}$
are easily expressed in terms of gamma functions.
This corresponds to the choice of the basis vectors in~\eqref{J1inv2}
and~\eqref{J2inv2}.

}

The Gram--Schmidt orthonormalisation procedure at the $n$-th step
determines $P_{\ell,n}(\eta;\bm{\lambda})$ by the following formula
\begin{gather*}
 P_{\ell,n}(\eta;\bm{\lambda})\propto
 \Xil{p_n(\eta)}-\sum_{m=0}^{n-1}P_{\ell,m}(\eta;\bm{\lambda})
 \bigl\langle P_{\ell,m}(\eta;\bm{\lambda}),\Xil{p_n(\eta)}
 \bigr\rangle_{\ell,\bm{\lambda}} h_{\ell,m}(\bm{\lambda})^{-1},
\end{gather*}
which is to be compared with the procedure for the undeformed ($\ell=0$)
polynomial
\begin{gather}
 P_n(\eta;\bm{\lambda'})\propto
 p_n(\eta)-\sum_{m=0}^{n-1}P_m(\eta;\bm{\lambda'})
 \bigl(P_m(\eta;\bm{\lambda'}),p_n(\eta)\bigr)_{\bm{\lambda'}}
 h_m(\bm{\lambda'})^{-1},
 \label{PnGS}
\end{gather}
where $\bm{\lambda'}=\bm{\lambda}+\ell\bm{\delta}+\bm{\tilde{\delta}}$.
These two orthonormalisations are essentially the same.
In fact, by applying $\Xil{\,\cdot\,}$ to \eqref{PnGS} and using
\eqref{Pln=Xil[Pn]}, we have
\begin{gather*}
 P_{\ell,n}(\eta;\bm{\lambda})\propto
 \Xil{p_n(\eta)}-\sum_{m=0}^{n-1}P_{\ell,m}(\eta;\bm{\lambda})
 d_0(m,\bm{\lambda})
 \bigl(P_m(\eta;\bm{\lambda'}),p_n(\eta)\bigr)_{\bm{\lambda'}}
 h_m(\bm{\lambda'})^{-1}.
\end{gather*}
Comparing these we obtain
\begin{gather*}
 \bigl\langle P_{\ell,m}(\eta;\bm{\lambda}),
 \Xil{p_n(\eta)}\bigr\rangle_{\ell,\bm{\lambda}}
  h_{\ell,m}(\bm{\lambda})^{-1}\nonumber\\
  \qquad{} =d_0(m,\bm{\lambda})
 \bigl(P_m(\eta;\bm{\lambda'}),p_n(\eta)\bigr)_{\bm{\lambda'}}
 h_m(\bm{\lambda'})^{-1},
 \qquad m=0,\ldots,n-1.
\end{gather*}
In other words, the Gram--Schmidt orthonormalisation for the undeformed
($\ell=0$) polynomials
\begin{gather*}
 p_n(\eta)\to P_n(\eta;\bm{\lambda'}),
\end{gather*}
provides that for the $X_{\ell}$ polynomials, too.

\section{Generating functions}\label{section9}

Generating functions for orthogonal polynomials have played another
important role in classical analysis.
Let us def\/ine generating functions for the $X_{\ell}$ polynomials
$P_{\ell,n}$ and for the undeformed polynomials $P_n$,
\begin{gather*}
 G_{\ell}(t,\eta;\bm{\lambda})\eqdef\sum_{n=0}^{\infty}
 t^nP_{\ell,n}(\eta;\bm{\lambda}),\qquad
 G(t,\eta;\bm{\lambda})\eqdef G_0(t,\eta;\bm{\lambda})
 =\sum_{n=0}^{\infty}t^nP_n(\eta;\bm{\lambda}).
\end{gather*}
The latter is quite well known
\begin{gather*}
 G(t,\eta;\bm{\lambda})=
 \begin{cases}
 G^{(g-\frac12)}(t,\eta),\quad & (\text{L})\\
 G^{(g-\frac12,h-\frac12)}(t,\eta),\quad & (\text{J})
 \end{cases}
\end{gather*}
where $G^{(\alpha)}(t,x)$ and $G^{(\alpha,\beta)}(t,x)$ are given in
\eqref{Lgenfn} and \eqref{Jgenfn}.
Since $P_{\ell,n}(\eta;\bm{\lambda})$ is expressed linearly in terms of
$P_n(\eta;\bm{\lambda'})$ \eqref{Plnform}
($\bm{\lambda'}=\bm{\lambda}+\ell\bm{\delta}+\bm{\tilde{\delta}}$),
the generating function $G_{\ell}(t,\eta;\bm{\lambda})$ is expressed
simply in terms of the known $G(t,\eta;\bm{\lambda'})$:
\begin{gather*}
 d_0(t\partial_t,\bm{\lambda})G_{\ell}(t,\eta;\bm{\lambda})
 =\sum_{n=0}^{\infty}t^nd_0(n,\bm{\lambda})P_{\ell,n}(\eta;\bm{\lambda})\nonumber\\
\phantom{d_0(t\partial_t,\bm{\lambda})G_{\ell}(t,\eta;\bm{\lambda})}{}
=d_1(\bm{\lambda})\xi_{\ell}(\eta;\bm{\lambda}+\bm{\delta})
 G(t,\eta;\bm{\lambda'})-d_2(\eta)\xi_{\ell}(\eta;\bm{\lambda})
  \partial_{\eta}G(t,\eta;\bm{\lambda'}).
\end{gather*}
The forward shift relation \eqref{Pnforward} implies
\begin{gather*}
 \partial_{\eta}G(t,\eta;\bm{\lambda})
 =\cF^{-1}t\,f_{t\,\partial_t+1}(\bm{\lambda})
 G(t,\eta;\bm{\lambda}+\bm{\delta}).
\end{gather*}

Here we present the concrete forms of the generating functions:
\begin{gather*}
 d_0(t\partial_t,\bm{\lambda})G_{\ell}(t,\eta;\bm{\lambda})\\
 \qquad{}=
 \begin{cases}
 \bigl(L_{\ell}^{(g+\ell-\frac12)}(-\eta)
 +\tfrac{t}{1-t}L_{\ell}^{(g+\ell-\frac32)}(-\eta)\bigr)
 G^{(g+\ell-\frac32)}(t,\eta),
\quad & (\text{L1})\vspace{2mm}\\[4pt]
 \bigl((g+\tfrac12)L_{\ell}^{(-g-\ell-\frac32)}(\eta)
 -\tfrac{t\eta}{1-t}L_{\ell}^{(-g-\ell-\frac12)}(\eta)\bigr)
 G^{(g+\ell+\frac12)}(t,\eta),
\quad & (\text{L2}) \vspace{2mm}\\
 \Bigl((h+\tfrac12)P_{\ell}^{(g+\ell-\frac12,-h-\ell-\frac32)}(\eta)
 +\tfrac{(1+\eta)t}{R}\bigl(\tfrac{1}{R}+\tfrac{g+\ell-\frac32}{1+R-t}
 +\tfrac{h+\ell+\frac12}{1+R+t}\bigr)
  & \vspace{2mm}\\
 \qquad{}\times P_{\ell}^{(g+\ell-\frac32,-h-\ell-\frac12)}(\eta)\Bigr)
 G^{(g+\ell-\frac32,h+\ell+\frac12)}(t,\eta),
\quad & (\text{J1})\vspace{2mm}\\
 \Bigl((g+\tfrac12)P_{\ell}^{(-g-\ell-\frac32,h+\ell-\frac12)}(\eta)
 -\tfrac{(1-\eta)t}{R}\bigl(\tfrac{1}{R}+\tfrac{g+\ell+\frac12}{1+R-t}
 +\tfrac{h+\ell-\frac32}{1+R+t}\bigr)
  & \vspace{2mm}\\
 \qquad{} \times P_{\ell}^{(-g-\ell-\frac12,h+\ell-\frac32)}(\eta)\Bigr)
 G^{(g+\ell+\frac12,h+\ell-\frac32)}(t,\eta), \quad & (\text{J2})
 \end{cases}
 \nonumber
\end{gather*}
where $R\eqdef\sqrt{1-2\eta t+t^2}$.

Next let us introduce the double generating function, that is the generating
function of the generating functions $G_{\ell}(t,\eta;\bm{\lambda})$:
\begin{gather*}
\mathcal{G}(s,t,\eta;\bm{\lambda})\eqdef
\sum_{\ell=0}^{\infty}s^{\ell}G_{\ell}(t,\eta;\bm{\lambda}).
\end{gather*}
For L1 and L2 cases, the explicit forms are:
\begin{alignat*}{3}
& (\text{L1}):\quad&&\mathcal{G}(s,t,\eta;g)
=2^{g-\frac32}\bigl((2-t)\sqrt{1-t}+t\sqrt{1-t-4s} \bigr) & \nonumber\\
&&& \phantom{\mathcal{G}(s,t,\eta;g)=}{} \times
\frac{e^{-\frac{t}{1-t}\eta
+\frac14s^{-1}\big(\sqrt{1-t}-\sqrt{1-t-4s} \big)^2\eta}}
{\sqrt{1-t-4s}{\sqrt{1-t}}^{g+\frac32}
\bigl(\sqrt{1-t}+\sqrt{1-t-4s}\,\bigr)^{g-\frac12}}, &
\\
& (\text{L2}):\quad& &d_0(t\partial_t,g)\mathcal{G}(s,t,\eta;g)
=\Bigl(g+\tfrac12-\frac{t(1-t+s)}{(1-t)^2}\eta\Bigr)
\frac{e^{-\frac{s+t}{1-t}\eta}}{(1-t+s)^{g+\frac32}},&
\end{alignat*}
which are obtained by using  the two shifted generating functions
\eqref{Lgenfn+} and \eqref{Lgenfn-} in Appendix~\ref{appendixE.1}. It
is a good challenge to derive the double generating functions for
the $X_\ell$ Jacobi polynomials.

\section{Three term recurrence relations}\label{section10}

Three term recurrence relations are one of the most fundamental
characteristics of the ordinary orthogonal polynomials of one variable.
Obviously the exceptional orthogonal polynomials do not satisfy these
relations. Nevertheless, being deformations of ordinary orthogonal polynomials,
the $X_{\ell}$ polynomials are expected to retain certain reminiscent
properties of the three term recurrence.

Here we present a simple modif\/ication of the three term recurrence
relations valid for the $X_{\ell}$ polynomials. Its relevance is,
however, as yet unclear. Let us denote the three term recurrence
relation for the Laguerre or the Jacobi polynomials as
\begin{gather*}
 \eta P_n(\eta;\bm{\lambda})
 =A_n(\bm{\lambda})P_{n+1}(\eta;\bm{\lambda})
 +B_n(\bm{\lambda})P_n(\eta;\bm{\lambda})
 +C_n(\bm{\lambda})P_{n-1}(\eta;\bm{\lambda}),
\end{gather*}
in which the explicit forms of the coef\/f\/icients $A_n$, $B_n$ and $C_n$
can be read from \eqref{L3term} and \eqref{J3term} in
Appendix~\ref{appendixE.1} and Appendix~\ref{appendixE.2}.
As a substitute of the above three term recurrence relations, we expect
that a certain element in the degree $\ell+n+1$ invariant polynomial
subspace $\mathcal{V}_{\ell,n+1}$ \eqref{Linv}, \eqref{Jinv}, which is
related to $\eta P_{\ell,n}(\eta;\bm{\lambda'})$, to be expressed in
terms of $P_{\ell,n+1}(\eta;\bm{\lambda})$,
$P_{\ell,n}(\eta;\bm{\lambda})$ and $P_{\ell,n-1}(\eta;\bm{\lambda})$.
{}From \eqref{Pln=Xil[Pn]} this can be achieved as
\begin{gather*}
 \Xil{\eta P_n(\eta;\bm{\lambda'})}
  =A_n(\bm{\lambda'})d_0(n+1,\bm{\lambda})P_{\ell,n+1}(\eta;\bm{\lambda})
 +B_n(\bm{\lambda'})d_0(n,\bm{\lambda})P_{\ell,n}(\eta;\bm{\lambda})\nonumber\\
\phantom{\Xil{\eta P_n(\eta;\bm{\lambda'})}=}{}
 +C_n(\bm{\lambda'})d_0(n-1,\bm{\lambda})P_{\ell,n-1}(\eta;\bm{\lambda}),
\end{gather*}
where $\bm{\lambda'}=\bm{\lambda}+\ell\bm{\delta}+\bm{\tilde{\delta}}$.

\section[Zeros of $X_{\ell}$ polynomials]{Zeros of $\boldsymbol{X_{\ell}}$ polynomials}\label{section11}

The zeros of orthogonal polynomials have always attracted the interest of
researchers. In the case of $X_{\ell}$ polynomials
$P_{\ell,n}(\eta;\bm{\lambda})$, it has $n$ zeros in the domain where
the weight function is def\/ined, that is $(0,\infty)$ for the L1 and L2
polynomials and $(-1,1)$ for the J1 and J2 polynomials.
See  for example, Section~5.4 of~\cite{askey}.
The behaviour of these zeros are the same as those of other ordinary
orthogonal polynomials.
This is guaranteed by the oscillation theorem of the one-dimensional
quantum mechanics, since $P_{\ell,n}(\eta;\bm{\lambda})$ are obtained
as the polynomial part of the eigenfunctions of a shape invariant
quantum mechanical problem~\cite{os16,os19}.

Here we discuss the location of the extra $\ell$ zeros of the exceptional
orthogonal polynomials,
which lie in various dif\/ferent positions for the
dif\/ferent types of polynomials. So far we have verif\/ied by direct
calculation for lower $\ell$ and~$n$:
The $\ell$ extra zeros of L1 polynomials are on the negative real line
$(-\infty,0)$.
Those of the L2 $X_{\ell \text{:odd}}$ polynomials are 1 real negative
zero which lies to the left of the remaining $\frac12(\ell-1)$  pairs of
complex conjugate roots.
The L2 $X_{\ell \text{:even}}$ polynomials have $\frac12\ell$ pairs of
complex conjugate roots.
For~L2, these $\ell$ additional roots lie to the left of the $n$ real zeros.

The situations for the $X_\ell$ Jacobi polynomials are a bit more complicated.
The J1 $X_{\ell \text{:odd}}$ polynomials have 1 real negative root
which lies to the left of the remaining $\frac12(\ell-1)$  pairs of complex
conjugate roots with negative  real parts.
The J1 $X_{\ell \text{:even}}$ polynomials have $\frac12\ell$ pairs of
complex conjugate roots with negative  real parts.
For~J1, some of the complex roots have real parts between~$-1$ and~0.
The J2 $X_{\ell \text{:odd}}$ polynomials have 1 real positive root
which lies to the right of the remaining $\frac12(\ell-1)$  pairs of complex
conjugate roots with positive  real parts.
The~J2 $X_{\ell \text{:even}}$ polynomials have $\frac12\ell$ pairs of
complex conjugate roots with positive  real parts.
For~J2, some of the complex roots have real parts between~0 and~1.

\section{Summary and comments}\label{section12}

We have given an in-depth study of the properties of the exceptional
($X_{\ell}$) polynomials discovered recently in \cite{os16, os19, os18}.
Our main focus is the derivation of certain equivalent but much simpler
looking forms of the $X_{\ell}$ polynomials. The derivation is based on
the analysis of the second order dif\/ferential equations for the $X_{\ell}$
polynomials within the framework of the Fuchsian dif\/ferential equations
in the entire complex plane. These new forms of the $X_{\ell}$ polynomials
allow easy verif\/ication of the actions of the forward and backward shift
operators on the $X_{\ell}$ polynomials, and provide direct derivation
of the Rodrigues formulas and the generating functions. The structure of
the invariant polynomial subspaces under the Fuchsian dif\/ferential
operators is elucidated.
The bases of the invariant polynomial subspaces provide a simple
substitute of the three term recurrence relations. The Gram--Schmidt
construction of the $X_{\ell}$ polynomials starting from the above
bases is demonstrated with the help of an integration formula.
The properties of the extra zeros of the $X_{\ell}$ polynomials are
discussed.
Some technical details are relegated to the Appendices.
The proof of the equivalence of the new and original forms of the
$X_{\ell}$ polynomials is given. Simple proofs of the forward and
backward shift operations are shown. The integration formula is
elementarily proven. Various fundamental formulas of the Laguerre
and Jacobi polynomials are supplied for easy reference.

Let us mention that the same method, deformation in terms of a
degree $\ell$ eigenpolynomial, applied to the discrete quantum
mechanical Hamiltonians for the Wilson and Askey--Wilson
polynomials produced two sets of inf\/initely many shape invariant
systems together with exceptional ($X_{\ell}$) Wilson and
Askey--Wilson polynomials ($\ell=1,2,\ldots$) \cite{os17}. It will
be interesting to carry out the same analysis in these discrete
cases.

Finally, concerning the issue of global solutions of Fuchsian
dif\/ferential equations, we would like to make a comment on the
well-known theorem by Heine--Stieltjes \cite{heine-st}. It asserts
the existence of a polynomial solution for the dif\/ferential
equation
\begin{gather*}
A(x)\frac{d^2}{dx^2}y(x)+B(x)\frac{d}{dx}y(x)+C(x)y(x)=0,
\end{gather*}
in  which only two coef\/f\/icient functions $A(x)$ and $B(x)$ are
specif\/ied. They are degree $p+1$ and~$p$ polynomials,
respectively. In this case a polynomial~$C(x)$ of degree~$p-1$ is
not given at the beginning but is determined so that the equation
admits a degree $n$ polynomial solution. Thus the problem setting
is not a proper Fuchsian dif\/ferential equation and the process of
determi\-ning~$C(x)$ and the polynomial solution is purely
algebraic. We consider these polynomial solutions do not qualify
to be `global solutions' of the ordinary Fuchsian equations.

After this paper was arXived, some of the results were re-derived
in terms of the
Darboux--Crum transformations \cite{gkm10,stz}.

\appendix

\section[Equivalence of different forms of $X_{\ell}$ polynomials]{Equivalence of dif\/ferent forms of $\boldsymbol{X_{\ell}}$ polynomials}\label{appendixA}

The exceptional Jacobi polynomial for the trigonometric DPT
presented in \cite{os16}, namely J2 exceptional polynomial, is
\begin{gather*}
 P_{\ell,n}(\eta;\bm{\lambda}) \eqdef
 \biggl(\xi_{\ell}(\eta;g+1,h+1)
 +\frac{2n(-g+h+\ell-1)\,\xi_{\ell-1}(\eta;g,h+2)}
 {(-g+h+2\ell-2)(g+h+2n+2\ell-1)}\nonumber\\
 \hphantom{P_{\ell,n}(\eta;\bm{\lambda}) \eqdef}{}
 -\frac{n(2h+4\ell-3)\,\xi_{\ell-2}(\eta;g+1,h+3)}
 {(2g+2n+1)(-g+h+2\ell-2)}\biggr)P_n(\eta;\bm{\lambda}+\ell\bm{\delta})\nonumber\\
\hphantom{P_{\ell,n}(\eta;\bm{\lambda}) \eqdef}{}
  +\frac{(-g+h+\ell-1)(2g+2n+2\ell-1)}{(2g+2n+1)(g+h+2n+2\ell-1)}
 \xi_{\ell-1}(\eta;g,h+2)P_{n-1}(\eta;\bm{\lambda}+\ell\bm{\delta}).
\end{gather*}
In this paper we have presented it in a much simpler form in
\eqref{XJform} (J2)
\begin{gather*}
  P_{\ell,n}(\eta;\bm{\lambda})=\frac{1}{n+g+\frac12}\Bigl(
 (g+\tfrac12)\xi_{\ell}(\eta;g+1,h+1)P_n(\eta;g+\ell+1,h+\ell-1)\nonumber\\
\phantom{P_{\ell,n}(\eta;\bm{\lambda})=\frac{-1}{n+g+\frac12}\Bigl(}
 -(1-\eta)\xi_{\ell}(\eta;g,h)
 \partial_{\eta}P_n(\eta;g+\ell+1,h+\ell-1)\Bigr).
\end{gather*}
In the following we write them as
$P_{\ell,n}^{\text{org}}(\eta;\bm{\lambda})$ and
$P_{\ell,n}^{\text{new}}(\eta;\bm{\lambda})$, respectively and show
that they are in fact equal by using various identities of the Jacobi
polynomials.
For lower $\ell$ and $n$ the equality can be verif\/ied by direct calculation.

We f\/ix $\ell$ and use new parameters $\alpha$ and $\beta$ instead of
$g$ and $h$,
\begin{gather*}
 \alpha\eqdef g+\ell-\tfrac12,\qquad \beta\eqdef h+\ell-\tfrac12.
\end{gather*}
By using the forward shift relation for the Jacobi polynomial~\eqref{Jforward}, the polynomials~$\xi_{\ell}$ and~$P_{\ell,n}$ are
expressed as
\begin{gather}
 \xi_{\ell}(\eta;\bm{\lambda})
 =P_{\ell}^{(-\alpha-1,\beta-1)}(\eta),
 \qquad \xi_{\ell-1}(\eta;\bm{\lambda})
 =P_{\ell-1}^{(-\alpha, \beta-2)}(\eta),
 \nonumber\\
  \xi_{\ell-2}(\eta;\bm{\lambda})
 =P_{\ell-2}^{(-\alpha+1,\,\beta-3)}(\eta),
 \nonumber\\
 P_{\ell,n}^{\text{org}}(\eta;\bm{\lambda}) =
 \Biggl(P_{\ell}^{(-\alpha-2,\beta)}(\eta)
 +\frac{2n(\ell-\alpha+\beta-1)P_{\ell-1}^{(-\alpha,\beta)}(\eta)}
 {(2\ell-\alpha+\beta-2)(2n+\alpha+\beta)}\nonumber\\
\qquad{}
 -\frac{n(\beta+\ell-1)P_{\ell-2}^{(-\alpha,\beta)}(\eta)}
 {(\alpha+n-\ell+1)(2\ell-\alpha+\beta-2)}\Biggr)
 P_n^{(\alpha,\beta)}(\eta)\nonumber\\
\qquad{}
  +\frac{(\ell-\alpha+\beta-1)(\alpha+n)}
 {(\alpha+n-\ell+1)(2n+\alpha+\beta)}
 P_{\ell-1}^{(-\alpha,\beta)}(\eta)P_{n-1}^{(\alpha,\beta)}(\eta),
 \label{os16PlnJac2}\\
 P_{\ell,n}^{\text{new}}(\eta;\bm{\lambda})
 =\frac{1}{\alpha+n-\ell+1}\Bigl(
 (\alpha-\ell+1)P_{\ell}^{(-\alpha-2,\beta)}(\eta)
 P_n^{(\alpha+1,\beta-1)}(\eta)\nonumber\\
\qquad{}
 -\tfrac12(n+\alpha+\beta+1)(1-\eta)P_{\ell}^{(-\alpha-1,\beta-1)}(\eta)
 P_{n-1}^{(\alpha+2,\beta)}(\eta)\Bigr).
 \label{PlnJac2}
\end{gather}
Here we provide the proof for the equivalence of the two expressions
\eqref{os16PlnJac2} and \eqref{PlnJac2} step by step:
\begin{gather*}
P_{\ell,n}^{\text{org}}(\eta;\bm{\lambda})
\stackrel{\text{(\romannumeral1)}}{=}
 \Biggl(
 \frac{(\ell+\beta)(1-\eta)P_{\ell-1}^{(-\alpha,\beta)}(\eta)
 +(\alpha+1)(1+\eta)P_{\ell-1}^{(-\alpha-1,\beta+1)}(\eta)}{-2\ell}\nonumber\\
 \qquad{}
 +\frac{2n(\ell-\alpha+\beta-1)P_{\ell-1}^{(-\alpha,\beta)}(\eta)}
 {(2\ell-\alpha+\beta-2)(2n+\alpha+\beta)}
 -\frac{n(\beta+\ell-1)}{(\alpha+n-\ell+1)(2\ell-\alpha+\beta-2)}\nonumber\\
 \qquad{}\times
 \frac{2(\ell-1)P_{\ell-1}^{(-\alpha,\beta)}(\eta)
 -(-\alpha+\beta+2\ell-2)(1+\eta)\bigl(P_{\ell-1}^{(-\alpha,\beta)}(\eta)
 -P_{\ell-1}^{(-\alpha-1,\beta+1)}(\eta)\bigr)}{-2(\beta+\ell-1)}
 \Biggr)\nonumber\\
 \qquad{}  \times
 P_n^{(\alpha,\beta)}(\eta)   +\frac{(\ell-\alpha+\beta-1)(\alpha+n)}
 {(\alpha+n-\ell+1)(2n+\alpha+\beta)}
 P_{\ell-1}^{(-\alpha,\beta)}(\eta)\nonumber\\
\qquad{} \times
 \frac{2nP_n^{(\alpha,\beta)}(\eta)-(\alpha+\beta+2n)(1-\eta)\bigl(
 P_n^{(\alpha,\beta)}(\eta)-P_n^{(\alpha+1,\beta-1)}(\eta)\bigr)}
 {2(\alpha+n)}\nonumber\\
 =\frac{1}{\alpha+n-\ell+1}\Biggl(
 \tfrac12(\ell-\alpha+\beta-1)(1-\eta)P_{\ell-1}^{(-\alpha,\beta)}(\eta)
 P_n^{(\alpha+1,\beta-1)}(\eta)\nonumber\\
\qquad{}
 +\frac{n+\alpha+1}{2\ell}\bigl(
 -(\alpha-\ell+1)(1+\eta)P_{\ell-1}^{(-\alpha-1,\beta+1)}(\eta)
 -\beta(1-\eta)P_{\ell-1}^{(-\alpha,\beta)}(\eta)\bigr)
 P_n^{(\alpha,\beta)}(\eta)\Biggr)\nonumber\\
 =\frac{1}{\alpha+n-\ell+1}\Biggl(
 (\alpha-\ell+1)\frac{(\ell+\beta)(1-\eta)P_{\ell-1}^{(-\alpha,\beta)}(\eta)
 +(\alpha+1)(1+\eta)P_{\ell-1}^{(-\alpha-1,\beta+1)}(\eta)}{-2\ell}\nonumber\\
\qquad{}
 \times P_n^{(\alpha+1,\beta-1)}(\eta)
 +\frac{(\ell-\alpha-1)(1+\eta)P_{\ell-1}^{(-\alpha-1,\beta+1)}(\eta)
 -\beta(1-\eta)P_{\ell-1}^{(-\alpha,\beta)}(\eta)}{2\ell}\nonumber\\
\qquad{}
 \times\bigl((n+\alpha+1)P_n^{(\alpha,\beta)}(\eta)
 -(\alpha+1)P_n^{(\alpha+1,\beta-1)}(\eta)\bigr)\Biggr)\nonumber\\
 \stackrel{\text{(\romannumeral2)}}{=}\frac{1}{\alpha+n-\ell+1}\Bigl(
 (\alpha-\ell+1)P_{\ell}^{(-\alpha-2,\beta)}(\eta)
 P_n^{(\alpha+1,\beta-1)}(\eta)\nonumber\\
\qquad{}
 +P_{\ell}^{(-\alpha-1,\beta-1)}(\eta)
 \bigl((n+\alpha+1)P_n^{(\alpha,\beta)}(\eta)
 -(\alpha+1)P_n^{(\alpha+1,\beta-1)}(\eta)\bigr)\Bigr)\nonumber\\
 \stackrel{\text{(\romannumeral3)}}{=}\frac{1}{\alpha+n-\ell+1}\Bigl(
 (\alpha-\ell+1)P_{\ell}^{(-\alpha-2,\beta)}(\eta)
 P_n^{(\alpha+1,\beta-1)}(\eta)\nonumber\\
\qquad{}
 +P_{\ell}^{(-\alpha-1,\beta-1)}(\eta)
 \tfrac{-1}{2}(n+\alpha+\beta+1)(1-\eta)
 P_{n-1}^{(\alpha+2,\beta)}(\eta)\Bigr)
 =P_{\ell,n}^{\text{new}}(\eta;\bm{\lambda}),
\end{gather*}
where we have used \eqref{Jid2}, \eqref{Jid4m} and \eqref{Jid4} in
(\romannumeral1), \eqref{Jid2} and \eqref{Jid2m} in (\romannumeral2),
and \eqref{Jid1m} in (\romannumeral3).

The other exceptional polynomials, the J1 case is obtained from the
above J2 case by \eqref{Jparity}. The equality of the exceptional
Laguerre polynomials given in \cite{os16,os19} and those given in this
paper \eqref{XLform} will not be given here, since the L1 and L2 cases
are obtained from the J1 and J2 cases by the limit~\eqref{J->L}.

\section{Forward and backward shift relations}\label{appendixB}

Here we provide proofs for the forward \eqref{FonP} and backward~\eqref{BonP} shift relations which apply equally for the four
types of $X_{\ell}$ polynomials. The method is elementary based on
various identities of the Laguerre and Jacobi polynomials.

The forward shift relation \eqref{FonP} is equivalent to a polynomial
identity \eqref{for_rel}.
By using $d_0(n-1,\bm{\lambda}+\bm{\delta})=d_0(n,\bm{\lambda})$,
$f_n(\bm{\lambda}+\bm{\tilde{\delta}})=f_n(\bm{\lambda})$ and~\eqref{Pnforward}, it is easy to show that r.h.s.\ of~\eqref{for_rel} can
be factorised
$\partial_{\eta}P_n(\eta;\bm{\lambda}+\ell\bm{\delta}+\bm{\tilde{\delta}})
\times(\cdots)$.
The forward shift relation is thus equivalent to $(\cdots)=0$, namely,
\begin{gather}
 0 =d_1(\bm{\lambda})\xi_{\ell}(\eta;\bm{\lambda}+\bm{\delta})^2
 -d_1(\bm{\lambda}+\bm{\delta})\xi_{\ell}(\eta;\bm{\lambda})
 \xi_{\ell}(\eta;\bm{\lambda}+2\bm{\delta})
 -\partial_{\eta}d_2(\eta)\xi_{\ell}(\eta;\bm{\lambda})
 \xi_{\ell}(\eta;\bm{\lambda}+\bm{\delta})\nonumber\\
\phantom{0=}{}
 +d_2(\eta)\xi_{\ell}(\eta;\bm{\lambda})
 \partial_{\eta}\xi_{\ell}(\eta;\bm{\lambda}+\bm{\delta})
 -d_2(\eta)\partial_{\eta}\xi_{\ell}(\eta;\bm{\lambda})
 \xi_{\ell}(\eta;\bm{\lambda}+\bm{\delta}).
 \label{fid4}
\end{gather}
This is a polynomial identity of degree $2\ell$ and it is quadratic
in the Laguerre/Jacobi polynomials.
This identity can be proven elementarily by using
$d_1(\bm{\lambda})-d_1(\bm{\lambda}+\bm{\delta})=\partial_{\eta}d_2(\eta)$
and \eqref{xil(l+d)} to eliminate
$\xi_{\ell}(\eta;\bm{\lambda}+2\bm{\delta})$ and
$\partial_{\eta}\xi_{\ell}(\eta;\bm{\lambda})$.

The backward shift relation \eqref{BonP} is equivalent to a polynomial
identity \eqref{back_rel}.
By using $d_0(n-1,\bm{\lambda}+\bm{\delta})=d_0(n,\bm{\lambda})$ and
$b_{n-1}(\bm{\lambda})=-2n$, the r.h.s.\ of~\eqref{back_rel} becomes
\begin{gather*}
 P_n(\eta;\bm{\lambda'})\times(\cdots)
 +\partial_{\eta}P_n(\eta;\bm{\lambda'})\times(\cdots)+P_{n-1}(\eta;\bm{\lambda'}+\bm{\delta})\times(\cdots)\nonumber\\
\qquad{}
 +\partial_{\eta}P_{n-1}(\eta;\bm{\lambda'}+\bm{\delta})\times(\cdots)
 +\partial_{\eta}^2P_{n-1}(\eta;\bm{\lambda'}+\bm{\delta})\times(\cdots),
 \nonumber
\end{gather*}
where $\bm{\lambda'}=\bm{\lambda}+\ell\bm{\delta}+\bm{\tilde{\delta}}$.
The above expression can be reduced to
$P_{n-1}(\eta;\bm{\lambda'}+\bm{\delta})\times X(\eta)
+\partial_{\eta}P_{n-1}(\eta;\bm{\lambda'}+\bm{\delta})
\times c_2(\eta)Y(\eta)$,
by using \eqref{Pnforward}, \eqref{Pnbackward}, \eqref{Pndiffeq} and
the relations
$(c_1(\eta;\bm{\lambda}+\bm{\delta}+\bm{\tilde{\delta}})
-c_1(\eta;\bm{\lambda}))d_2(\eta)=2c_2(\eta)\partial_{\eta}d_2(\eta)$
with $\mathcal{E}_n(\bm{\lambda})=f_n(\bm{\lambda})b_{n-1}(\bm{\lambda})$.
Up to an overall normalization, this $Y(\eta)$ is just the r.h.s.\ of~\eqref{fid4}, so it vanishes.
Hence the backward shift relation is equivalent to $X(\eta)=0$, namely,
\begin{gather}
 0 =c_1(\eta,\bm{\lambda'})d_1(\bm{\lambda})
 \xi_{\ell}(\eta;\bm{\lambda}+\bm{\delta})^2
 -c_1(\eta,\bm{\lambda}+\ell\bm{\delta})d_1(\bm{\lambda}+\bm{\delta})
 \xi_{\ell}(\eta;\bm{\lambda})\xi_{\ell}(\eta;\bm{\lambda}+2\bm{\delta})\nonumber\\
\phantom{0=}{}
 -c_2(\eta)d_1(\bm{\lambda}+\bm{\delta})
 \bigl(\xi_{\ell}(\eta;\bm{\lambda})
 \partial_{\eta}\xi_{\ell}(\eta;\bm{\lambda}+2\bm{\delta})
 -\partial_{\eta}\xi_{\ell}(\eta;\bm{\lambda})
 \xi_{\ell}(\eta;\bm{\lambda}+2\bm{\delta})\bigr)\nonumber\\
\phantom{0=}{}
 +\tfrac14\mathcal{E}_1(\bm{\lambda'})d_2(\eta)
 \xi_{\ell}(\eta;\bm{\lambda})\xi_{\ell}(\eta;\bm{\lambda}+\bm{\delta}).
 \label{bid4}
\end{gather}
This is a polynomial identity of degree $2\ell+1$ and it is quadratic
in the Laguerre/Jacobi polynomials.
With the help of \eqref{xil(l+d)} and \eqref{xildiffeq}, it is elementary
to show that the r.h.s.\ of~\eqref{bid4} becomes
$\xi_{\ell}(\eta;\bm{\lambda}+\bm{\delta})d_1(\bm{\lambda})d_2(\eta)
\times(\cdots)$.
Up to an overall normalization, this $(\cdots)$ part is just
(l.h.s.\ of~\eqref{xil(l)})~$-$ (r.h.s.\ of~\eqref{xil(l)}), so it vanishes.
This concludes the proof of the backward shift relation.

\section{Proof of the integration formula}\label{appendixC}

Here we present a simple proof for the integration formulas
\eqref{innerform} or \eqref{intformula}. We evaluate
\begin{gather*}
 \bigl\langle\Xil{p(\eta)},\Xil{q(\eta)}\bigr\rangle_{\ell,\bm{\lambda}}=\int
 \bigl(d_1(\bm{\lambda})\xi_{\ell}(\eta;\bm{\lambda}+\bm{\delta})p(\eta)
 -d_2(\eta)\xi_{\ell}(\eta;\bm{\lambda})\partial_{\eta}p(\eta)\bigr)\nonumber\\
\qquad{}\times
 \bigl(d_1(\bm{\lambda})\xi_{\ell}(\eta;\bm{\lambda}+\bm{\delta})q(\eta)
 -d_2(\eta)\xi_{\ell}(\eta;\bm{\lambda})\partial_{\eta}q(\eta)\bigr)
 \frac{W(\eta;\bm{\lambda}+\ell\bm{\delta})}
 {\xi_{\ell}(\eta;\bm{\lambda})^2}d\eta\nonumber\\
=\int\Biggl(
 d_2(\eta)^2 \partial_{\eta}p(\eta)\partial_{\eta}q(\eta)
 +\left(
 \frac{d_1(\bm{\lambda})\xi_{\ell}(\eta;\bm{\lambda}+\bm{\delta})}
 {\xi_{\ell}(\eta;\bm{\lambda})}\right)^2
 p(\eta)q(\eta)\nonumber\\
\qquad{}
 -\frac{d_1(\bm{\lambda})\xi_{\ell}(\eta;\bm{\lambda}+\bm{\delta})}
 {\xi_{\ell}(\eta;\bm{\lambda})}d_2(\eta)
 \partial_{\eta}\bigl(p(\eta)q(\eta)\bigr)\Biggr)
 W(\eta;\bm{\lambda}+\ell\bm{\delta})d\eta\nonumber\\
\stackrel{\text{(\romannumeral1)}}{=}
 \int W(\eta;\bm{\lambda}+\ell\bm{\delta})\Biggl(
 d_2(\eta)^2\,\partial_{\eta}p(\eta)\partial_{\eta}q(\eta)\nonumber\\
\qquad
 +\left(\left(
 \frac{d_1(\bm{\lambda})\xi_{\ell}(\eta;\bm{\lambda}+\bm{\delta})}
 {\xi_{\ell}(\eta;\bm{\lambda})}\right)^2
 - \frac{\partial_{\eta}\bigl(W(\eta;\bm{\lambda}+\ell\bm{\delta})
 \frac{d_1(\bm{\lambda})\xi_{\ell}(\eta;\bm{\lambda}+\bm{\delta})}
 {\xi_{\ell}(\eta;\bm{\lambda})}d_2(\eta)\bigr)}
 {W(\eta;\bm{\lambda}+\ell\bm{\delta})}
 \right)
 p(\eta)q(\eta)\Biggr)d\eta \nonumber\\
\stackrel{\text{(\romannumeral2)}}{=}
 \int\Bigl(d_1(\bm{\lambda})d_3(\bm{\lambda}+\ell\bm{\delta},\ell)
 p(\eta)q(\eta)W(\eta;\bm{\lambda'})
 +\tfrac14\cF^2\partial_{\eta}p(\eta)\partial_{\eta}q(\eta)
 W(\eta;\bm{\lambda'}+\bm{\delta})
 \Bigr)d\eta,
\end{gather*}
where $\bm{\lambda'}=\bm{\lambda}+\ell\bm{\delta}+\bm{\tilde{\delta}}$.
In (\romannumeral1) we have integrated by part and in (\romannumeral2)
we have used
\begin{gather*}
 \left(
 \frac{d_1(\bm{\lambda})\xi_{\ell}(\eta;\bm{\lambda}+\bm{\delta})}
 {\xi_{\ell}(\eta;\bm{\lambda})}\right)^2
 -\frac{\partial_{\eta}\bigl(W(\eta;\bm{\lambda}+\ell\bm{\delta})
 \frac{d_1(\bm{\lambda})\xi_{\ell}(\eta;\bm{\lambda}+\bm{\delta})}
 {\xi_{\ell}(\eta;\bm{\lambda})}d_2(\eta)\bigr)}
 {W(\eta;\bm{\lambda}+\ell\bm{\delta})}\nonumber\\
 \qquad{}
  =d_1(\bm{\lambda})d_3(\bm{\lambda}+\ell\bm{\delta},\ell)
 \frac{d_2(\eta)^2}{c_2(\eta)},
\end{gather*}
which is shown by using \eqref{xildiffeq} and \eqref{xil(l+d)} to eliminate
$\xi_{\ell}(\eta;\bm{\lambda}+\bm{\delta})$.

\section{Hyperbolic DPT potential}\label{appendixD}

Here we provide a brief summary of the properties of the $X_\ell$ Jacobi
polynomials related to the {\em deformed hyperbolic\/} DPT potential.
They are of the J2 type.
The $\ell=1$ case was introduced in \cite{BQR} and the general $\ell$ case was
studied in \cite{os16}.
In contrast to the radial oscillator and the trigonometric DPT potentials,
the undeformed hyperbolic DPT potential allows only a f\/inite number of
square integrable polynomials
$P_n(\eta,\bm{\lambda})$, $n=0,1,\ldots,n_B\eqdef[\tfrac12(h-g)]'$,
where $[x]'$ denotes the greatest integer not equal or exceeding $x$.
The situation is the same for the exceptional polynomials,
$P_{\ell,n}(\eta,\bm{\lambda})$, $n=0,1,\ldots,n_B-\ell$.
Except for this point, the arguments for the hyperbolic DPT are the same as
the radial oscillator and the trigonometric DPT potential cases.
In particular, the much simpler looking new forms of the polynomials are
also equivalent to the original forms of the polynomials given in~\cite{os16}.
We present various data:
\begin{gather*}
 \bm{\lambda}\eqdef(g,h),\quad h>g>0,\qquad\bm{\delta}\eqdef(1,-1),\qquad
P_n(\eta;g,h)\eqdef P_n^{(g-\frac12,-h-\frac12)}(\eta),
 \nonumber\\
  \xi_{\ell}(\eta;g,h)\eqdef
 P_{\ell}^{(-g-\ell-\frac12,-h+\ell-\frac32)}(\eta),\qquad
 n\leq n_B,\quad \ell<n_B,\\
P_{\ell,n}(\eta;\bm{\lambda}) \quad \eqref{Plnform},\qquad
 n=0,1,\ldots,n_B-\ell,\\
 d_0(n,\bm{\lambda})\eqdef n+g+\tfrac12,\qquad
 \bm{\tilde{\delta}}\eqdef(1,1),\qquad
 d_1(\bm{\lambda})\eqdef g+\tfrac12,\qquad
 d_2(\eta)\eqdef 1-\eta,\\
 W(\eta;\bm{\lambda})\eqdef\tfrac{1}{2^{g-h+1}}
 (\eta-1)^{g-\frac12}(\eta+1)^{-h-\frac12},\qquad
 1<\eta<\infty,\\
 h_n(\bm{\lambda})\eqdef\frac{\Gamma(n+g+\frac12)\Gamma(h-g-n+1)}
 {2\,n!\,(h-g-2n)\Gamma(h-n+\frac12)},\\
 h_{\ell,n}(g,h)\eqdef h_n(g+\ell,h-\ell)
 \frac{(n+g+\ell+\frac12)(h-n-2\ell+\frac12)}
 {(n+g+\frac12)(h-n-\ell+\frac12)},\\
 c_1(\eta,\bm{\lambda})\eqdef g+h+(g-h+1)\eta,\qquad
 c_2(\eta)\eqdef\eta^2-1,\\
 \tilde{c}_1(\eta,\bm{\lambda},\ell)\eqdef h-g-2\ell+1-(g+h)\eta,\\
 \mathcal{E}_n(\bm{\lambda})\eqdef4n(h-g-n),\qquad
 \widetilde{\mathcal{E}}_{\ell}(\bm{\lambda})\eqdef
 4\ell(g+h+1-\ell),\\
 d_3(\bm{\lambda},\ell)\eqdef h-\ell+\tfrac12,\qquad
 \cF\eqdef 4,\qquad
 f_n(\bm{\lambda})\eqdef 2(n+g-h),\qquad
 b_{n-1}(\bm{\lambda})\eqdef-2n,\\
 P_{\ell,n}(\eta;\bm{\lambda})
 =\frac{1}{2^nn!}
 \frac{\xi_{\ell}(\eta;\bm{\lambda})}
 {(\eta-1)^{g+\ell-\frac12}(\eta+1)^{-h+\ell-\frac12}}
 \prod_{k=0}^{n-1}\Bigl(\frac{d}{d\eta}+\partial_{\eta}\log
 \frac{\xi_{\ell}(\eta;\bm{\lambda}+(k+1)\bm{\delta})}
 {\xi_{\ell}(\eta;\bm{\lambda}+k\bm{\delta})}\Bigr)\nonumber\\
 \phantom{P_{\ell,n}(\eta;\bm{\lambda})=}\times
 \frac{(\eta-1)^{n+g+\ell-\frac12}(\eta+1)^{n-h+\ell-\frac12}}
 {\xi_{\ell}(\eta;\bm{\lambda}+n\bm{\delta})}
\xi_{\ell}\bigl(\eta;\bm{\lambda}+(n+1)\bm{\delta}\bigr),\\
 d_0(t\partial_t,\bm{\lambda})G_{\ell}(t,\eta;\bm{\lambda})
 = \sum\limits_n
 t^nd_0(n,\bm{\lambda})P_{\ell,n}(\eta;\bm{\lambda})\nonumber\\
 =\Bigl((g+\tfrac12)P_{\ell}^{(-g-\ell-\frac32,-h+\ell-\frac12)}(\eta)
 -\tfrac{(1-\eta)t}{R}\bigl(\tfrac{1}{R}+\tfrac{g+\ell+\frac12}{1+R-t}
 +\tfrac{h-\ell+\frac12}{1+R+t}\bigr)
 P_{\ell}^{(-g-\ell-\frac12,-h+\ell-\frac32)}(\eta)\Bigr)\nonumber\\
 \qquad{}\times
 G^{(g+\ell+\frac12,h-\ell+\frac12)}(t,\eta),\qquad
 R\eqdef\sqrt{1-2\eta t+t^2},\\
 \mathcal{V}_{\ell,n}\eqdef
 \text{Span}\bigl[\eta^k(g+\tfrac12)\xi_{\ell}(\eta;g+1,h-1)
 -k(1-\eta)\eta^{k-1}\xi_{\ell}(\eta;g,h);k=0,1,\ldots,n\bigr],\\
 p_n(\eta)=(\eta-1)^n\quad\text{for \eqref{GSpn}}.
\end{gather*}
In the expression of the generating function, the summation range is
extended to inf\/inity, $\sum\limits_{n=0}^{n_B-\ell}\to\sum\limits_{n=0}^{\infty}$,
formally.
The forward and backward shift relations \eqref{for_rel} and
\eqref{back_rel} in terms of the Jacobi polynomials are the same as
the J2 case.

\section{Summary: properties of the polynomials}\label{appendixE}

\subsection{Laguerre polynomials}\label{appendixE.1}

\noindent$\bullet$ Def\/inition (expansion formula)
\begin{gather}
 L_n^{(\alpha)}(x)=\frac{1}{n!}
 \sum_{k=0}^n\frac{(-n)_k}{k!}(\alpha+k+1)_{n-k}x^k.
 \label{Lexp}
\end{gather}

\noindent$\bullet$ Forward and backward shift relations
\begin{gather}
  \partial_xL_n^{(\alpha)}(x)=-L_{n-1}^{(\alpha+1)}(x),
 \label{Lforward}\\
 x\partial_xL_{n-1}^{(\alpha+1)}(x)+(\alpha+1-x)L_{n-1}^{(\alpha+1)}(x)
 =nL_{n}^{(\alpha)}(x).
 \label{Lbackward}
\end{gather}

\noindent$\bullet$ Dif\/ferential equation
\begin{gather}
 x\partial_x^2L_n^{(\alpha)}(x)
 +(\alpha+1-x)\partial_xL_n^{(\alpha)}(x)
 +nL_n^{(\alpha)}(x)=0.
 \label{Ldiffeq}
\end{gather}

\noindent$\bullet$ Three term recurrence relation
\begin{gather}
 xL_n^{(\alpha)}(x)=-(n+1)L_{n+1}^{(\alpha)}(x)
 +(2n+\alpha+1)L_n^{(\alpha)}(x)
 -(n+\alpha)L_{n-1}^{(\alpha)}(x).
 \label{L3term}
\end{gather}

\noindent$\bullet$ Rodrigues formula
\begin{gather}
 L_n^{(\alpha)}(x)=\frac{1}{n!}\frac{1}{e^{-x}x^{\alpha}}
 \left(\frac{d}{dx}\right)^n\bigl(e^{-x}x^{n+\alpha}\bigr).
 \label{LRod}
\end{gather}

\noindent$\bullet$ Orthogonality ($\alpha>-1$)
\begin{gather}
 \int_0^{\infty}dx\,e^{-x}x^{\alpha}L_n^{(\alpha)}(x)L_m^{(\alpha)}(x)
 =\frac{1}{n!}\,\Gamma(n+\alpha+1)\delta_{nm}.
 \label{Lortho}
\end{gather}

\noindent$\bullet$ Generating functions
\begin{gather}
G^{(\alpha)}(t,x) \eqdef\sum_{n=0}^{\infty}t^nL_n^{(\alpha)}(x)
 =\frac{e^{-\frac{tx}{1-t}}}{(1-t)^{\alpha+1}},
 \label{Lgenfn}\\
G^{(\alpha)}_+(t,x) \eqdef\sum_{n=0}^{\infty}t^nL_n^{(\alpha+n)}(x)
=\frac{2^{\alpha}e^{-\frac14t^{-1}(1-\sqrt{1-4t}\,)^2x}}
{\sqrt{1-4t}\,\bigl(1+\sqrt{1-4t}\,\bigr)^{\alpha}},
\label{Lgenfn+}\\
G^{(\alpha)}_-(t,x) \eqdef\sum_{n=0}^{\infty}t^nL_n^{(\alpha-n)}(x)
=(1+t)^{\alpha}e^{-tx}.
\label{Lgenfn-}
\end{gather}

Formulas \eqref{Lgenfn+} and \eqref{Lgenfn-} can be derived by
using \eqref{Lforward} and \eqref{Ldiffeq}.

\noindent$\bullet$ Identities
\begin{gather}
 L_n^{(\alpha)}(x)-L_n^{(\alpha-1)}(x)=L_{n-1}^{(\alpha)}(x),
 \label{Lid1}\\
 xL_{n-1}^{(\alpha+1)}(x)-\alpha L_{n-1}^{(\alpha)}(x)
 =-nL_n^{(\alpha-1)}(x),
 \label{Lid2}
\end{gather}
which can be verif\/ied elementarily based on \eqref{Lexp}.

\subsection{Jacobi polynomials}\label{appendixE.2}

\noindent$\bullet$ Def\/inition (expansion formula)
\begin{gather}
 P_n^{(\alpha,\beta)}(x)=\frac{(\alpha+1)_n}{n!}
 \sum_{k=0}^n\frac{1}{k!}\frac{(-n)_k(n+\alpha+\beta+1)_k}{(\alpha+1)_k}
 \left(\frac{1-x}{2}\right)^k.
 \label{Jexp}
\end{gather}

\noindent$\bullet$ Parity
\begin{gather}
 P_n^{(\alpha,\beta)}(-x)=(-1)^nP_n^{(\beta,\alpha)}(x).
 \label{Jparity}
\end{gather}

\noindent$\bullet$ Forward and backward shift relations
\begin{gather}
  \partial_xP_n^{(\alpha,\beta)}(x)
 =\tfrac12(n+\alpha+\beta+1)P_{n-1}^{(\alpha+1,\beta+1)}(x),
 \label{Jforward}\\
  \big(1-x^2\big)\partial_xP_{n-1}^{(\alpha+1,\beta+1)}(x)
 +\bigl(\beta-\alpha-(\alpha+\beta+2)x\bigr)P_{n-1}^{(\alpha+1,\beta+1)}(x)
 =-2nP_{n}^{(\alpha,\beta)}(x).
 \label{Jbackward}
\end{gather}

\noindent$\bullet$ Dif\/ferential equation
\begin{gather}
 \big(1-x^2\big)\partial_x^2P_n^{(\alpha,\beta)}(x)
 +\bigl(\beta-\alpha-(\alpha+\beta+2)x\bigr)\partial_x P_n^{(\alpha,\beta)}(x)\nonumber\\
 \qquad{}
 +n(n+\alpha+\beta+1)P_n^{(\alpha,\beta)}(x)=0.
 \label{Jdiffeq}
\end{gather}

\noindent$\bullet$ Three term recurrence relation
\begin{gather}
 xP_n^{(\alpha,\beta)}(x) =
 \frac{2(n+1)(n+\alpha+\beta+1)P_{n+1}^{(\alpha,\beta)}(x)}
 {(2n+\alpha+\beta+1)(2n+\alpha+\beta+2)}
 +\frac{(\beta^2-\alpha^2)P_n^{(\alpha,\beta)}(x)}
 {(2n+\alpha+\beta)(2n+\alpha+\beta+2)}\nonumber\\
\phantom{xP_n^{(\alpha,\beta)}(x) =}{}
 +\frac{2(n+\alpha)(n+\beta)P_{n-1}^{(\alpha,\beta)}(x)}
 {(2n+\alpha+\beta)(2n+\alpha+\beta+1)}.
 \label{J3term}
\end{gather}

\noindent$\bullet$ Rodrigues formula
\begin{gather}
 P_n^{(\alpha,\beta)}(x)=\frac{(-1)^n}{2^nn!}
 \frac{1}{(1-x)^{\alpha}(1+x)^{\beta}}
 \left(\frac{d}{dx}\right)^n\bigl((1-x)^{n+\alpha}(1+x)^{n+\beta}\bigr).
 \label{JRod}
\end{gather}

\noindent$\bullet$ Orthogonality ($\alpha,\beta>-1$)
\begin{gather}
 \int_{-1}^1dx(1-x)^{\alpha}(1+x)^{\beta}
 P_n^{(\alpha,\beta)}(x)P_m^{(\alpha,\beta)}(x)\nonumber\\
 \qquad{}
 =\frac{2^{\alpha+\beta+1}}{n!}
 \frac{\Gamma(n+\alpha+1)\Gamma(n+\beta+1)}
 {(2n+\alpha+\beta+1)\Gamma(n+\alpha+\beta+1)}\delta_{nm}.
 \label{Jortho}
\end{gather}

\noindent$\bullet$ Generating function
\begin{gather} \label{Jgenfn}
 G^{(\alpha,\beta)}(t,x)\eqdef\sum_{n=0}^{\infty}t^nP_n^{(\alpha,\beta)}(x)
 =\frac{2^{\alpha+\beta}}{R(1+R-t)^{\alpha}(1+R+t)^{\beta}},\\
 R\eqdef\sqrt{1-2xt+t^2}.\nonumber
\end{gather}

\noindent$\bullet$ Identities
\begin{gather}
 2(n+\beta)P_n^{(\alpha,\beta-1)}(x)
 -2\beta P_n^{(\alpha-1,\beta)}(x)
 =(n+\alpha+\beta)(1+x)P_{n-1}^{(\alpha,\beta+1)}(x),
 \label{Jid1}\\
 (n+\beta)(1-x)P_{n-1}^{(\alpha+1,\beta)}(x)
 -\alpha(1+x)P_{n-1}^{(\alpha,\beta+1)}(x)
 =-2nP_n^{(\alpha-1,\beta)}(x),
 \label{Jid2}\\
 2(\alpha+n)P_{n-1}^{(\alpha,\beta)}(x)
 +(\alpha+\beta+2n)(1-x)\bigl(P_n^{(\alpha,\beta)}(x)
 -P_n^{(\alpha+1,\beta-1)}(x)\bigr)
 =2nP_n^{(\alpha,\beta)}(x),\!\!\!
 \label{Jid4}\\
 2(n+\alpha)P_n^{(\alpha-1,\beta)}(x)
 -2\alpha P_n^{(\alpha,\beta-1)}(x)
 =-(n+\alpha+\beta)(1-x)P_{n-1}^{(\alpha+1,\beta)}(x),
 \label{Jid1m}\\
 (n+\alpha)(1+x)P_{n-1}^{(\alpha,\beta+1)}(x)
 -\beta(1-x)P_{n-1}^{(\alpha+1,\beta)}(x)
 =2nP_n^{(\alpha,\beta-1)}(x),
 \label{Jid2m}\\
 -2(\beta\!+\!n)P_{n-1}^{(\alpha,\beta)}(x)
 +(\alpha\!+\!\beta\!+\!2n)(1+x)\bigl(P_n^{(\alpha,\beta)}(x)
  - P_n^{(\alpha{-}1,\beta{+}1)}(x)\bigr)
  = 2nP_n^{(\alpha,\beta)}(x). \!\!\!\!
 \label{Jid4m}
\end{gather}
Equations~\eqref{Jid1}--\eqref{Jid4} can be verif\/ied elementarily based on
\eqref{Jexp}. By using \eqref{Jparity}, equations~\eqref{Jid1m}--\eqref{Jid4m}
are obtained from equations~\eqref{Jid1}--\eqref{Jid4} respectively.

\noindent$\bullet$ Limit to the Laguerre polynomial
\begin{gather}
 \lim_{\beta\to\infty}P_n^{(\alpha,\pm\beta)}\bigl(1-\tfrac{2x}{\beta}\bigr)
 =L_n^{(\alpha)}(\pm x).
 \label{J->L}
\end{gather}
Various formulas for the Jacobi polynomials reduce to those for the Laguerre
polynomials in this limit.

\subsection*{Acknowledgments}

This work is supported in part by the National Science Council
(NSC) of the Republic of China under Grant NSC
96-2112-M-032-007-MY3 (CLH), and in part by Grants-in-Aid for
Scientif\/ic Research from the Ministry of Education, Culture,
Sports, Science and Technology, No.19540179 (RS). Part of the work
was done during CLH's visit to the Yukawa Institute for
Theoretical Physics (YITP), Kyoto University, and he would like to
thank the staf\/f and members of YITP for the hospitality extended
to him. RS wishes to thank National Taiwan University and National
Chiao-Tung University for the hospitality extended to him during
his visits in
which a part of the work was done.

\pdfbookmark[1]{References}{ref}
\LastPageEnding

\end{document}